\documentclass[prx,aps,twocolumn,english,superscriptaddress,longbibliography]{revtex4-2}
\usepackage[unicode=true,pdfusetitle, bookmarks=true,bookmarksnumbered=false,bookmarksopen=false, breaklinks=false,pdfborder={0 0 0},pdfborderstyle={},backref=false,colorlinks=true] {hyperref}

\usepackage{color,soul}
\usepackage{graphicx}
\usepackage{dcolumn}
\usepackage{bm}
\usepackage[mathlines]{lineno}
\usepackage{amsmath}

\begin{document}

\preprint{APS/123-QED}

\title{Superconducting Non-Reciprocity Based on Time-Modulated Coupled-Resonator Systems}

\author{Yi Zhuang}
\affiliation{Department of Electrical and Systems Engineering, Washington University in St.\ Louis, Missouri 63130}
\author{Chandrashekhar Gaikwad}
\affiliation{Department of Physics, Washington University, St.\ Louis, Missouri 63130}
\author{Daria Kowsari}
\affiliation{Department of Physics, Washington University, St.\ Louis, Missouri 63130}
\author{Kater Murch}
\affiliation{Department of Physics, Washington University, St.\ Louis, Missouri 63130}
\author{Aravind Nagulu}
\affiliation{Department of Electrical and Systems Engineering, Washington University in St.\ Louis, Missouri 63130}
\date{\today}
\begin{abstract}

We present a  unified approach for designing a diverse range of superconducting non-reciprocal components, including circulators, isolators, and uni-directional amplifiers, based on temporally-modulated coupled resonator networks. Our method leverages standard SQUID-based resonators as building blocks, arranged in various configurations such as series-coupled, wye-connected, and lattice-coupled resonators, to realize a wide range of on-chip non-reciprocal devices. Our theoretical studies demonstrated the effectiveness of the proposed approach, achieving circulators and isolators with near-zero insertion losses and isolation greater than 20 dB, and directional amplifiers with forward gain exceeding 10 dB and reverse isolation greater than 20 dB. To validate our findings, we implemented and measured a series-coupled three-resonator superconducting isolator using a single-layer superconducting process. At a base temperature of 20 mK, our device exhibited  insertion loss of 1.3 dB in the forward direction, and 
isolation of up to 25 dB at the center frequency and greater than 15 dB across a bandwidth of 250 MHz in the reverse direction. Our approach promises to enable the design of a broad range of high-performance non-reciprocal devices for superconducting circuits.

\end{abstract}

\maketitle



\section{Introduction}
\label{sec:Introduction}

Superconducting quantum systems are rapidly becoming a promising platform for building quantum computers and other quantum information processing devices~\cite{Arute_2019,Bardin_2020,Bardin_2021}. These systems consist of carefully engineered superconducting quantum bits (qubits) made using Josephson Junctions (JJs) or a parallel combination of JJs known as superconducting quantum interference devices (SQUIDs) and are operated at milliKelvin temperatures (10mK--100mK) to harness quantum effects~\cite{Josephson_1962,Rowell_1963}. Non-reciprocal components, such as circulators and isolators operating at these low temperatures, are widely used to protect the qubits from the noise and spurs of the downstream electronics at higher temperatures, and to separate the input and amplified signals in quantum-limited reflection-type amplifiers~\cite{Leo_MicrowMag2019,Nagulu_NatElectron_2020}. 
A typical qubit readout chain consists of 3--4 circulators per qubit. Currently, commercial ferrite circulators that violate Lorentz reciprocity when biased with a strong magnetic field (around 1 mT) are used for this purpose. These ferrite devices, however, cannot be integrated on-chip alongside the superconducting qubits due to the significant stray flux generated by their strong magnetic bias and the requirement of high deposition temperatures. As a result, they are implemented as connectorized microwave components with strong magnetic shielding, resulting in bulky form factors and expensive implementation costs. This poses challenges for their use in large-scale quantum computing systems with thousands of qubits in a single dilution refrigerator. 

Alternatively, Lorentz reciprocity can be broken using time-varying structures~\cite{Alu_ProcIEEE_2020,Leo_MicrowMag2019,Nagulu_NatElectron_2020} and has been explored extensively in various branches of physics ranging from acoustics~\cite{Alu_Science2014,Fbar_circ_Bhave,Matteo_MEMS2018}, electronics~\cite{kamal1960,Alu_NatPhys_2014,NRK_NatComm16,TD_NatComm17,Nagulu_NatElectron_2020,biedka2017ultra}, mechanics, and optics~\cite{Haldane_TI_2008,tzuang2014non,chamanara2017optical} for realizing magnet-free, on-chip non-reciprocal devices. Recognizing the need for miniaturized and monolithically integrated non-reciprocal devices in superconducting quantum systems, prior works have explored achieving on-chip non-reciprocal devices in superconducting platforms~\cite{Kerckhoff2015,Lehnert_PRX_2017,Aumentado_PRA_2017,Chow_2017,Ranzani_PRA_2017,Stace_PRL_2018,Chapman_PRA_2019,Bretheau_PRR_2021,Richman_PRXQuantum2021}. 
Additionally, a compact and scalable design of integrated superconducting non-reciprocal elements would also result in practical realizations of superconducting Floquet topological lattices similar to photonic~\cite{Fan_NatPhoton_2012}, acoustic~\cite{Fleury_TI_2016} and electronic~\cite{Nagulu_NatElectron_2022} Floquet topological insulators (TIs). Superconducting topological insulators could open the door to scalable and efficient-multiplexed qubit readout and control, arbitrary signal routing, and could enable potential applications that require integrated nonreciprocity in a lattice~\cite{Girvin_PRA_2010,Paetznick_PRXQuantum_2023}. 

In this work, we introduce a unified methodology to realize a wide gamut of superconducting non-reciprocal components such as isolators, circulators, and uni-directional amplifiers using the concept of temporally-modulated coupled resonator networks. The concept of temporal modulation in coupled resonators was introduced to realize on-chip circulators for wireless systems~\cite{Alu_NatPhys_2014, Alu_IMS_2017}. Later, this concept has been translated to realize isolating bandpass filters at radio frequencies (RF) where varactors were used as the modulating element~\cite{Alejandro_TMTT_2019,Gomes_TMTT_2019}. In this work, we present a methodology to translate this concept to superconducting circuits through inductance modulation by using SQUIDs as the modulating elements. Our study showed that it is possible to utilize standard SQUID-based resonators as fundamental components that can be arranged in different configurations, including series-coupled resonators, wye-connected resonators, and lattice-coupled resonators. These configurations allow for the implementation of a broad range of on-chip non-reciprocal responses, such as isolation, circulation, and directional amplification. Finally, we validate our method through the physical implementation and measurement of a series-coupled three-resonator superconducting isolator achieving $>$20 dB non-reciprocity in its amplitude response.

The rest of the article is structured as follows: section~\ref{sec:UnitCell} introduces the concept of time modulation and outlines the operation of the elementary unit-cell---a time-modulated SQUID-based resonator---which serves as the fundamental building block for our nonreciprocal devices. In section~\ref{sec:devices}, we delve into the concept, analytical studies, and simulation results of integrated non-reciprocal devices, including isolators, directional amplifiers, and circulators, that rely on coupled time-modulated resonators. Section~\ref{sec:implementation} presents the implementation and measurement results of a superconducting isolator constructed using three series-coupled SQUID-based resonators. Section~\ref{sec:FutureOutlook} discusses possible extensions of the proposed concept. Lastly, in section~\ref{sec:conclusion}, we conclude the paper by providing some final remarks.

\nopagebreak

\section{Time-Modulated Resonators}
\label{sec:UnitCell}
Consider a linear time-varying (LTV) system with one shunt component modulated with a periodic signal at a frequency $\omega_m$. The ABCD network properties of a parametrically modulated shunt element can be represented as
\begin{equation}
\label{eq:ABCD_time}
    \begin{bmatrix}
V_1(t) \\
I_1(t) 
\end{bmatrix} = \begin{bmatrix}
1 &0 \\
Y(t) & 1 
\end{bmatrix}  \times \begin{bmatrix} V_2(t)\\
I_2(t)\end{bmatrix},\end{equation}

\noindent where $V_{1,2}(t)$ and $I_{1,2}(t)$ are the voltages and currents at ports 1 and 2 respectively. By taking a Fourier transform of the system in \eqref{eq:ABCD_time}, one can show that the port voltages and currents carry the intermodulation signal between the input and the pump frequencies, namely, they contain frequencies component at $\left(\omega_{in}\pm k\omega_m \right)$ where $k=0,\pm1,\pm2,...$ In the spectral domain, such a time-modulated system can be represented as
\begin{equation}
    \begin{bmatrix}
\underline{V_1} \\
\\
\underline{I_1} 
\end{bmatrix} = \begin{bmatrix}
\underline{U} &\underline{0} \\
\\
\underline{Y} & \underline{U} 
\end{bmatrix} \times \begin{bmatrix} \underline{V_2}\\
\\
\underline{I_2}\end{bmatrix},\end{equation}

\noindent where $\underline{V_{1,2}}$ and $\underline{I_{1,2}}$ are column vectors of size $(2N+1)$ with Fourier coefficients of frequency components $(\omega_{in}\pm k\omega_m)$, $k= [-N, -(N-1),...0,... N-1, N]$, $\underline{Y}$ represents the spectral admittance matrix of the shunt element (see section \ref{sec:Methods} for more details), $\underline{U}$ and $\underline{0}$ are the identity and zero matrices~\cite{Desoer1959,Kurth1977}. The value of $N$ determines the accuracy of the spectral domain computation.

\subsection{Spectral Admittance Matrix of a DC-SQUID}


JJs are superconducting devices made by sandwiching a thin layer of insulator between two superconducting layers \cite{Josephson_1962,Rowell_1963}. A SQUID consists of two JJs in parallel and its inductance is controlled by modulating the magnetic flux threading the junction loop. The inductance of a SQUID can be expressed as 

\begin{equation}
\label{eq:SQUID}
L_\mathrm{SQUID} = \frac{\Phi_0}{ 4\pi I_c \cos(\pi\Phi/\Phi_0)},
\end{equation}
\noindent where $I_c$ is the critical current of the JJs, $\Phi$ is the magnetic flux threading the SQUID loop, and $\Phi_0$ is the flux quantum. It has to be noted that \eqref{eq:SQUID} is applicable for signal currents that are smaller than the critical current. Hence the power handling of the SQUID is limited by the value of critical current. The power handling of the SQUID can be increased by using an array of concatenated SQUID loops and the inductance of an $N$-stacked SQUID array multiplied by a factor of $N$. From \eqref{eq:SQUID}, the inductance of a flux-modulated SQUID can be expressed as

\begin{equation}
 \frac{1}{L(t)} = \frac{4\pi I_c}{\Phi_0} \cos\left(\frac{\pi(\Phi_{DC}+ \Delta\Phi\cos(\omega_mt+\theta))}{\Phi_0}\right).
 \label{eq:L_inverse}
\end{equation}
 
 The cosine term can be expanded and simplified using Taylor expansions in \eqref{eq:cos} and \eqref{eq:sin}. 
 \begin{table*}
\begin{equation}
\cos\left(\pi\frac{\Delta\phi cos(\omega_{m}t+\theta)}{\phi_0}\right) \approx
\left[1 - \frac{1}{4}\left(\frac{\pi\Delta\phi}{\phi_0}\right)^2 
\right]+\cos(2\omega_mt+2\theta)\left[- \frac{1}{4}\left(\frac{\pi\Delta\phi}{\phi_0}\right)^2 
\right]
\label{eq:cos}
\end{equation}

\begin{equation}
  \sin\left(\pi\frac{\Delta\phi cos(\omega_{m}t+\theta)}{\phi_0}\right) \approx \sin(\omega_mt+\theta)\left[\left(\frac{\pi\Delta\phi}{\phi_0}\right) - \frac{1}{8}\left(\frac{\pi\Delta\phi}{\phi_0}\right)^3\right] +\sin(3\omega_mt+3\theta)\left[\frac{1}{24}\left(\frac{\pi\Delta\phi}{\phi_0}\right)^3\right] \label{eq:sin}\end{equation}
\end{table*}
Further, the inductance of the SQUID can be approximated as

 \begin{equation}
\label{eq:L(t)}
     \frac{1}{L(t)}\approx\sum_{p=-3}^{3}F_{p}e^{jp\times \theta}e^{jp\omega_m t},
 \end{equation}

\noindent where the constants are

\begin{equation}\label{eq:F0}
F_0 = \frac{4\pi I_c }{\Phi_0} \cos\left(\frac{\pi\Phi_{DC}}{\Phi_0}\right)\left[1-\frac{1}{4}\left(\frac{\pi\Delta\Phi}{\Phi_0}\right)^2\right],\end{equation}
\begin{equation}
\begin{aligned}
F_{-1} = -F_1 &= \frac{2\pi I_c}{j\Phi_0} \sin\left(\frac{\pi\Phi_{DC}}{\Phi_0}\right)\times \\
&\quad\quad\left[\left(\frac{\pi\Delta\Phi}{\Phi_0}\right)-\frac{1}{8}\left(\frac{\pi\Delta\Phi}{\Phi_0}\right)^3\right], \label{eq:F1}
\end{aligned}
\end{equation}

\begin{equation}
F_{-2}=F_2 = \frac{2\pi I_c}{\phi_0}\cos\left(\frac{\pi\phi_{DC}}{\phi_0}\right)\left[-\frac{1}{4}\left(\frac{\pi\Delta\phi}{\phi_0}\right)^2\right],\end{equation}

\begin{equation}
F_{-3}= -F_3 = \frac{2\pi I_c}{\phi_0}\sin\left(\frac{\pi\phi_{DC}}{\phi_0}\right)\left[\frac{1}{24}(\frac{\pi\Delta\phi}{\phi_0})^3\right]. \end{equation}
The spectral representation of time-varying inductance is discussed in Appendix~\ref{sec:Methods}. Finally, from \eqref{eq:L(t)}, the spectral admittance of a flux-modulated SQUID can be expressed as \eqref{eq:spectral}.

\begin{table*}
\begin{equation}
\underline{Y_L} = \begin{bmatrix}
\ddots & \vdots & \vdots & \vdots & \vdots & \vdots & \reflectbox{$\ddots$} \\ \\

\cdots & \frac{F_{0}}{j(\omega-2\omega_m)} &  \frac{F_{-1}e^{-j\theta}}{j(\omega-\omega_m)} &  \frac{F_{-2}e^{-j2\theta}}{j\omega} & \frac{F_{-3}e^{-j3\theta}}{j(\omega+\omega_m)} & 0 &\cdots\\ \\

\cdots & \frac{F_{1}e^{j\theta}}{j(\omega-2\omega_m)} & \frac{F_{0}}{j(\omega-\omega_m)} &  \frac{F_{-1}e^{-j\theta}}{j\omega} &  \frac{F_{-2}e^{-j2\theta}}{j(\omega+\omega_m)} & \frac{F_{-3}e^{-j3\theta}}{j(\omega+2\omega_m)} &\cdots\\ \\

\cdots & \frac{F_{2}e^{j2\theta}}{j(\omega-2\omega_m)} & \frac{F_{1}e^{j\theta}}{j(\omega-\omega_m)} &  \frac{F_0}{j\omega} &  \frac{F_{-1}e^{-j\theta}}{j(\omega+\omega_m)} & \frac{F_{-2}e^{-j2\theta}}{j(\omega+2\omega_m)} &\cdots\\ \\

\cdots &\frac{F_{3}e^{j3\theta}}{j(\omega-2\omega_m)} & \frac{F_{2}e^{j2\theta}}{j(\omega-\omega_m)} & \frac{F_{1}e^{j\theta}}{j\omega} & \frac{F_0}{j(\omega +\omega_m)} & \frac{F_{-1}e^{-j\theta}}{j(\omega +2\omega_m)}  &\cdots\\ \\

\cdots & 0 & \frac{F_{-3}e^{-j3\theta}}{j(\omega-\omega_m)}& \frac{F_{-2}e^{-j2\theta}}{j\omega} & \frac{F_{-1}e^{-j\theta}}{j(\omega +\omega_m)} & \frac{F_0}{j(\omega +2\omega_m)} &\cdots\\ \\

\reflectbox{$\ddots$} & \vdots & \vdots & \vdots & \vdots & \vdots & \ddots 
\label{eq:spectral}
\end{bmatrix}
\end{equation}
\end{table*}

It is important to note that, the up-conversion and down-conversion frequency translational terms of a single time-modulated SQUID have reciprocal magnitude response and non-reciprocal phase response. 

\begin{figure*}[t]
\centering
\includegraphics[width=0.8\textwidth]{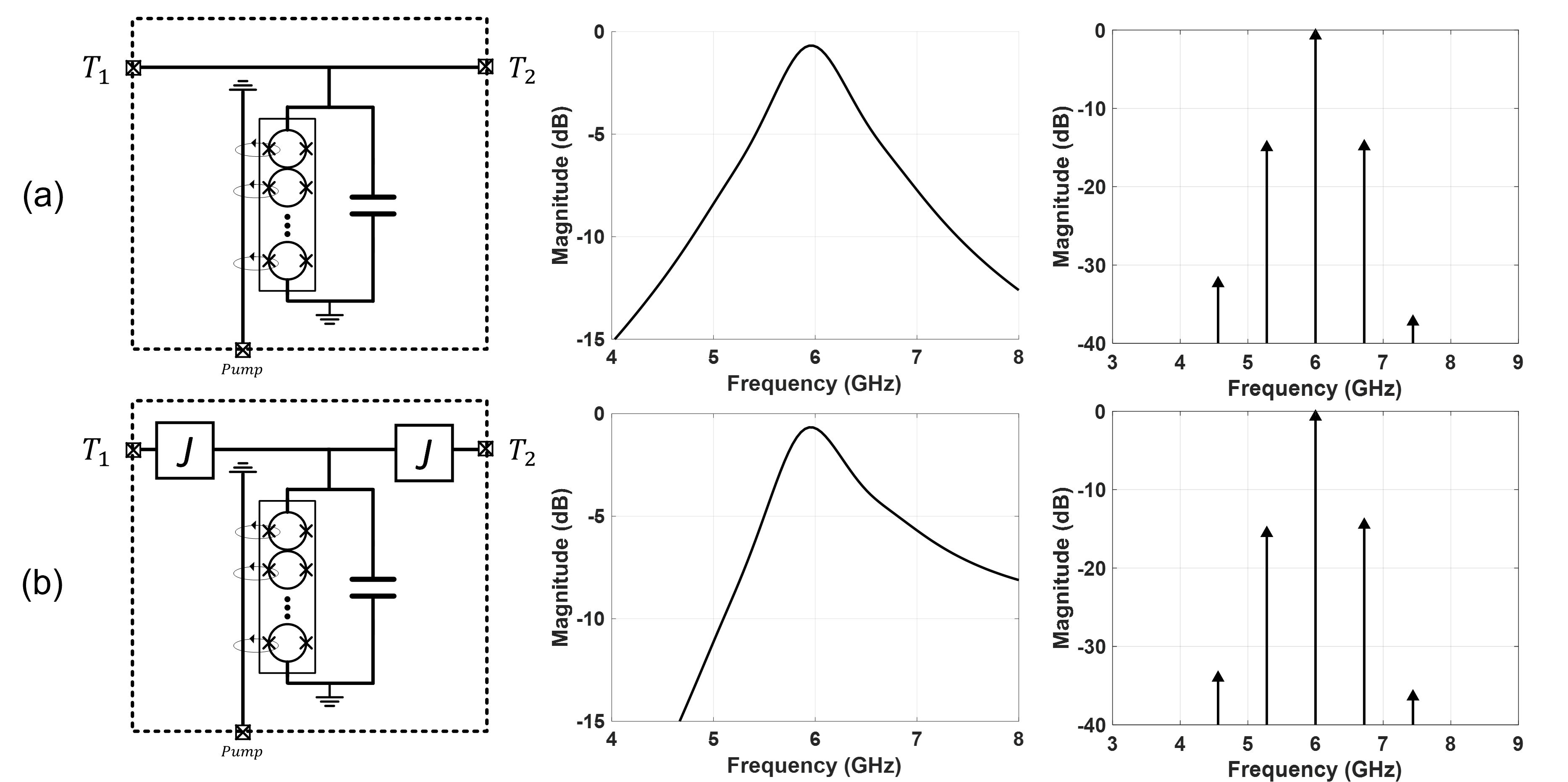}
\caption{\label{fig:concept} Schematic, transmission response, and normalized output spectra of a time modulated (a) unit cell realized using SQUID and a capacitor, and (b) unit cell sandwiched between two J-inverters. In both cases, the SQUID is biased at $\Phi_{DC}=$0.35$\Phi_0$, and modulated with $\Delta\Phi=$ 0.025$\Phi_0$, $f_m=$ 0.7~GHz, and an input signal is incident on port 1 at the center frequency, $f_{in}=f_{center}=6$ GHz.}
\end{figure*}

\subsection{Unit Cell: A Time-Modulated SQUID-based Resonator}
The building block of the time-modulated non-reciprocal components developed in this article is a resonator consisting of a flux-modulated SQUID in parallel with  a capacitor as shown in Fig.~\ref{fig:concept}(a). The ABCD matrix of the unit cell can be written as
\begin{equation}
\label{eq:LC_SQUID}
    \begin{bmatrix}
\underline{V_1} \\
\\
\underline{I_1} 
\end{bmatrix} = \begin{bmatrix}
\underline{U} &\underline{0} \\
\\
\underline{Y_{LC}} & \underline{U} 
\end{bmatrix} \times \begin{bmatrix} \underline{V_2}\\
\\
\underline{I_2}\end{bmatrix},\end{equation}
where $\underline{Y_{LC}}=\underline{Y_L}+\underline{Y_C}$, $\underline{Y_L}$ and $\underline{Y_C}$ is the spectral admittance matrix of the flux-modulated SQUID and capacitor. Since the capacitor is just a static element, $\underline{Y_C}$ reduces to a diagonal matrix with each entry corresponding to the admittance of the capacitance at the corresponding intermodulation (IM) frequency. Fig.~\ref{fig:concept}(a) also depicts the frequency response and normalized output spectra of a time-modulated resonator with $\Delta\Phi=$ 0.025$\Phi_0$, $f_m=$ 0.7~GHz, and an input signal at the center frequency $f_{in}=f_{center}=6$ GHz. The time-modulated resonators exhibit a reciprocal response with non-zero loss due to the frequency conversion of the input power to intermodulation frequencies. In section \ref{sec:devices}, we will show how multiple resonators can be coupled to create a direction-dependent frequency translation, thus realizing a non-reciprocal amplitude response. 

\subsection{Unit Cell with Admittance Inverter}
To enable coupling between resonator networks, we sandwich our unit cell between two admittance inverters (J-inverters) as shown in Fig. \ref{fig:concept}(b). Since the admittance inverters are time-invariant, the ABCD matrix is
\begin{equation}
M_J = \begin{bmatrix}
\underline{0} & \pm\frac{1}{jJ}\underline{U}\\
\pm jJ\underline{U} & \underline{0}. 
\end{bmatrix}, 
\label{eqn: M_J}
\end{equation} where 1/J is the characteristic impedance of the impedance inverter, and $j=\sqrt{-1}$. An appropriate value of the J-inverter admittance can be chosen to create the required coupling \cite{Pozar_textbook}. The spectral ABCD matrix of a unit cell with J-inverters can be expressed as
\begin{equation}
\begin{aligned}
    \begin{bmatrix}
\underline{V_1} \\
\\
\underline{I_1} 
\end{bmatrix} &= [M_{J}\times M_{LC}\times M_{J}] \times \begin{bmatrix} \underline{V_2}\\
\\
\underline{I_2}\end{bmatrix},\\
&= \begin{bmatrix}
\underline{U} &(\underline{Y_{LC}})/J^2\\
\\
\underline{0} & \underline{U} 
\end{bmatrix} \times \begin{bmatrix} \underline{V_2}\\
\\
\underline{I_2}\end{bmatrix}.\\
\end{aligned}
\end{equation}

As shown in Fig.~\ref{fig:concept}(b), the transmission response and the frequency translation features of the unit cell remain similar to  Fig.~\ref{fig:concept}(a). However, the J-inverters enable us to create coupling between two unit cells which is essential for creating a coupled resonator network and therefore non-reciprocal devices.

\begin{figure*}
\centering
\includegraphics[width=\textwidth]{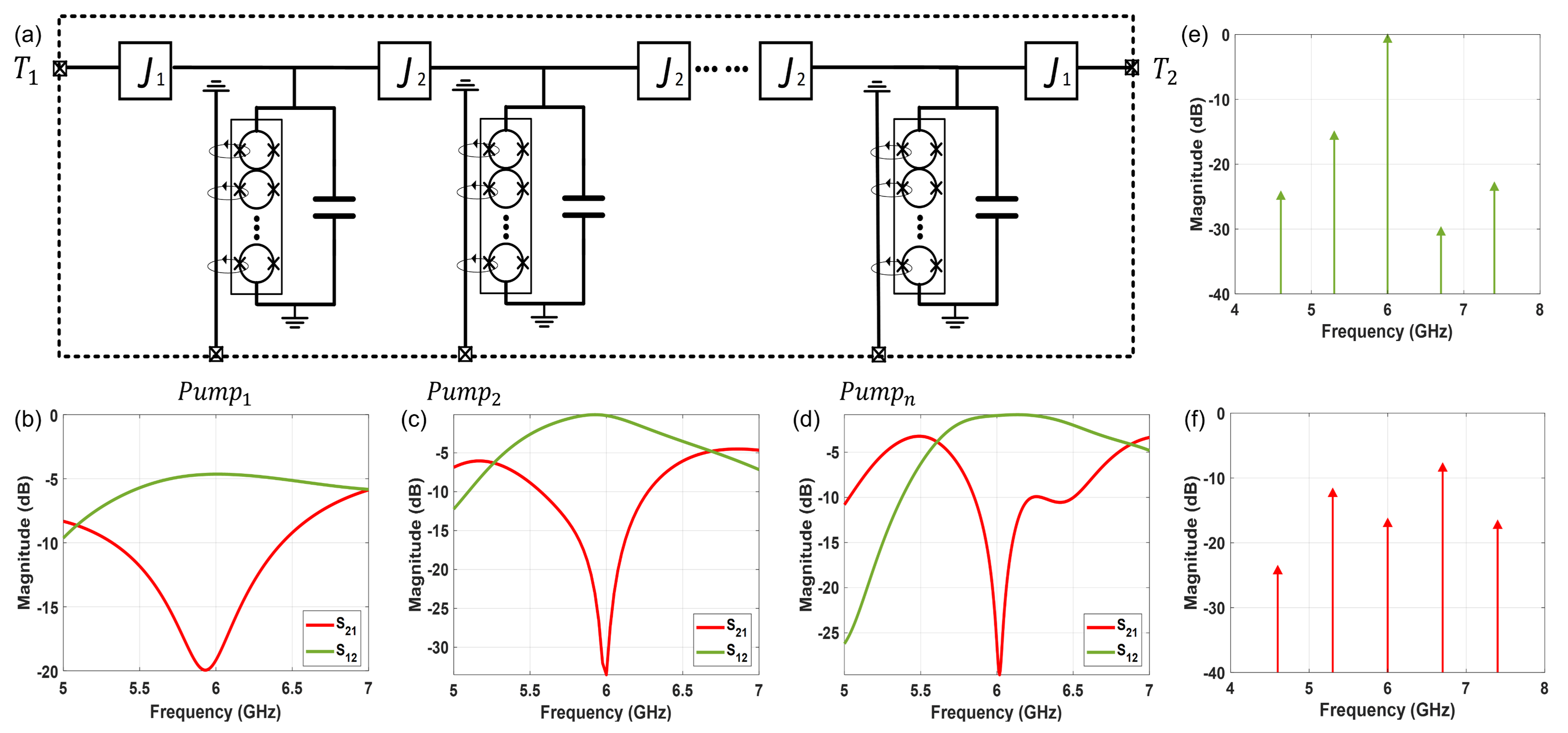}
\caption{\label{fig:bpf} (a) A general solution of building a SQUID-based higher-order bandpass filter. Response of a time-modulated non-reciprocal bandpass filter with (b) two coupled resonators ($\Phi_{DC}=0.35\Phi_0$, $\Delta\Phi=0.025\Phi_0$, $f_m=650~\mathrm{MHz}$, and $\theta=90^{\circ}$), (c) three coupled resonators ($\Phi_{DC}=0.35\Phi_0$, $\Delta\Phi=0.025\Phi_0$, $f_m=700~\mathrm{MHz}$, and $\theta=90^{\circ}$), and (d) four coupled resonators ($\Phi_{DC}=0.35\Phi_0$, $\Delta\Phi=0.025\Phi_0$, $f_m=700~\mathrm{MHz}$, and $\theta=90^{\circ}$). Simulated output spectra in (e) transmission and (f) isolation directions of a third-order non-reciprocal bandpass filter for an input excitation at the center frequency.}
\end{figure*} 

\section{Superconducting Non-Reciprocal Devices Based on Time-Modulated Coupled Resonator Networks}
\label{sec:devices}
In modern circuit design, the ability to achieve non-reciprocal response is essential. One approach to realize the such a response is by coupling several unit cells together using J-inverters, thus introducing a spatio-temporal modulation.  In addition, these temporally modulated unit cells can be connected in various circuit topologies such as series-coupling, wye-coupling, delta-coupling, and 2-D lattices \cite{Alu_NatPhys_2014,Fleury_TI_2016,Kord_TMTT2018,Gomes_TMTT_2019}. Each of these topologies can exhibit unique non-reciprocal behavior such as isolation, circulator, and topological robustness. In this section, we will discuss these circuit topologies and analyze their non-reciprocal behavior using the spectral-ABCD matrices. Specifically, we will investigate the factors that influence the non-reciprocal behavior, such as the modulation scheme and the arrangement of the unit cells. Understanding these dependencies allows us to design circuits with tailored non-reciprocal responses to meet specific application requirements. A similar theoretical study of time-modulated, coupled-resonator networks was reported in~\cite{Naaman_PRXQuantum2022}. 

\subsection{Amplitude Non-Reciprocity Using Coupled Unit Cells}
\label{sec:amp_NR}
When multiple time-varying elements are coupled, the input signal will be up-converted (down-converted) by one resonator and the generated IM products can be down-converted (up-converted) back to the input frequency by another resonator. From section \ref{sec:UnitCell}, we have seen that a single time-varying resonator results in the same conversion gain for the up-conversion and down-conversion to IM products, but results in a nonreciprocal phase relation. The nonreciprocal phase relation was not emphasized in section II. Therefore, if we introduce a phase staggering between the pumping signals of two time-varying resonators, the phase of the signal that gets reconstructed back to the input frequency would depend on the phase difference between the pumping signals. Under optimal modulation conditions the reconstructed input signal can add-up destructively or constructively depending on the incident signal direction, thus resulting in non-reciprocal amplitude response.

This scenario can be illustrated using a simple system with two unit cells that are coupled with one J-inverter and are modulated with $\Phi_{pump1}=\Delta\Phi\cos(\omega_mt+\theta_1)$ and $\Phi_{pump2}=\Delta\Phi\cos(\omega_mt+\theta_2)$. For the sake of simplicity and intuitive understanding, only the first IM conversion is considered in this illustration. However, for a more precise performance evaluation, conversions to other IM frequencies should also be taken into account, as done in the later sections. The network ABCD parameters of this system can be expressed as

\begin{equation}
    \begin{aligned}
    \begin{bmatrix}
\underline{V_1} \\
\\
\underline{I_1} 
\end{bmatrix}& = [M_{LC,1}\times M_{J}\times M_{LC,2}] \times \begin{bmatrix} \underline{V_2}\\
\\
\underline{I_2}\end{bmatrix},\\
& = \begin{bmatrix}
\frac{1}{jJ}\underline{Y_{LC,2}} &\frac{1}{jJ}\underline{U}\\
\\
jJ\underline{U}+\frac{1}{jJ}\underline{Y_{LC,1}}.\underline{Y_{LC,2}} & \frac{1}{jJ}\underline{Y_{LC,1}} 
\end{bmatrix} \times \begin{bmatrix} \underline{V_2}\\
\\
\underline{I_2}\end{bmatrix},
\end{aligned}
\end{equation}

\noindent where $M_{LC,1}$ and $M_{LC,2}$ are the spectral-ABCD matrices of the first and second resonators, $M_{J}$ is the spectral ABCD matrix of the coupling J-inverter, and 

\begin{equation} 
\underline{Y_{LC,1}} =\begin{bmatrix}
\ddots&\vdots &  \vdots &  \vdots&\reflectbox{$\ddots$} \\ \\
  \cdots&\frac{F_0-(\omega-\omega_m)^2C}{j(\omega-\omega_m)} &  \frac{F_{-1}e^{-j\theta_1}}{j\omega} &  \frac{F_{-2}e^{-j2\theta_1}}{j(\omega+\omega_m)}&\cdots \\ \\
  \cdots&\frac{F_1e^{j\theta_1}}{j(\omega-\omega_m)} &  \frac{F_0-\omega^2C}{j\omega} &  \frac{F_{-1}e^{-j\theta_1}}{j(\omega+\omega_m)}&\cdots \\ \\
 \cdots&\frac{F_2e^{j2\theta_1}}{j(\omega-\omega_m)} & \frac{F_1e^{j\theta_1}}{j\omega} & \frac{F_0-(\omega+\omega_m)^2C}{j(\omega +\omega_m)}&\cdots \\
 \reflectbox{$\ddots$}&\vdots &  \vdots &  \vdots& \ddots
\label{eq:Y_L}
\end{bmatrix},
\end{equation}

\noindent where $F_0$, $F_{-1}$, and $F_1$ are expressed in \eqref{eq:F0} and \eqref{eq:F1}. Similarly, $\underline{Y_{LC,2}}$ can be expressed in terms of $\theta_2$.
The transmission scattering parameter of this network can be expressed as

\begin{equation}
\begin{aligned}
\underline{S_{21}} &= 2[\underline{A}+\underline{B}/Z_o+\underline{C}Z_o+\underline{D}]^{-1}\\
&= \frac{j}{J}[\underline{Y_{LC,2}}+\underline{Y_{LC,1}}+\frac{\underline{U}}{Z_o}-J^2Z_o\underline{U}\\
&\qquad+Z_o\underline{Y_{LC,1}}.\underline{Y_{LC,2}}]^{-1},
\end{aligned}  
\end{equation}

 \noindent and by symmetry, one can show that
\begin{equation}
\begin{aligned}
\underline{S_{12}} &= \frac{j}{J}[\underline{Y_{LC,1}}+\underline{Y_{LC,2}}+\frac{\underline{U}}{Z_o}-J^2Z_o\underline{U}\\
&\qquad+Z_o\underline{Y_{LC,2}}.\underline{Y_{LC,1}}]^{-1}.
\end{aligned}  
\end{equation}

The eigenvalues of $\underline{Y_{LC,1}}$ and $\underline{Y_{LC,2}}$ are different when $\theta_1\neq\theta_2$, thereby making these matrices non-commuting  (i.e., $\underline{Y_{LC,2}}.\underline{Y_{LC,1}}\neq \underline{Y_{LC,1}}.\underline{Y_{LC,2}}$), resulting in nonreciprocal scattering matrices. By choosing an appropriate modulation amplitude, modulation frequency, and phase difference, one can design for $S_{21}\approx 1$ and $S_{12}\approx 0$ for the input and output frequency of $\omega$, thus resulting in an isolator.

\begin{figure*}
\centering
\includegraphics[width=\textwidth]{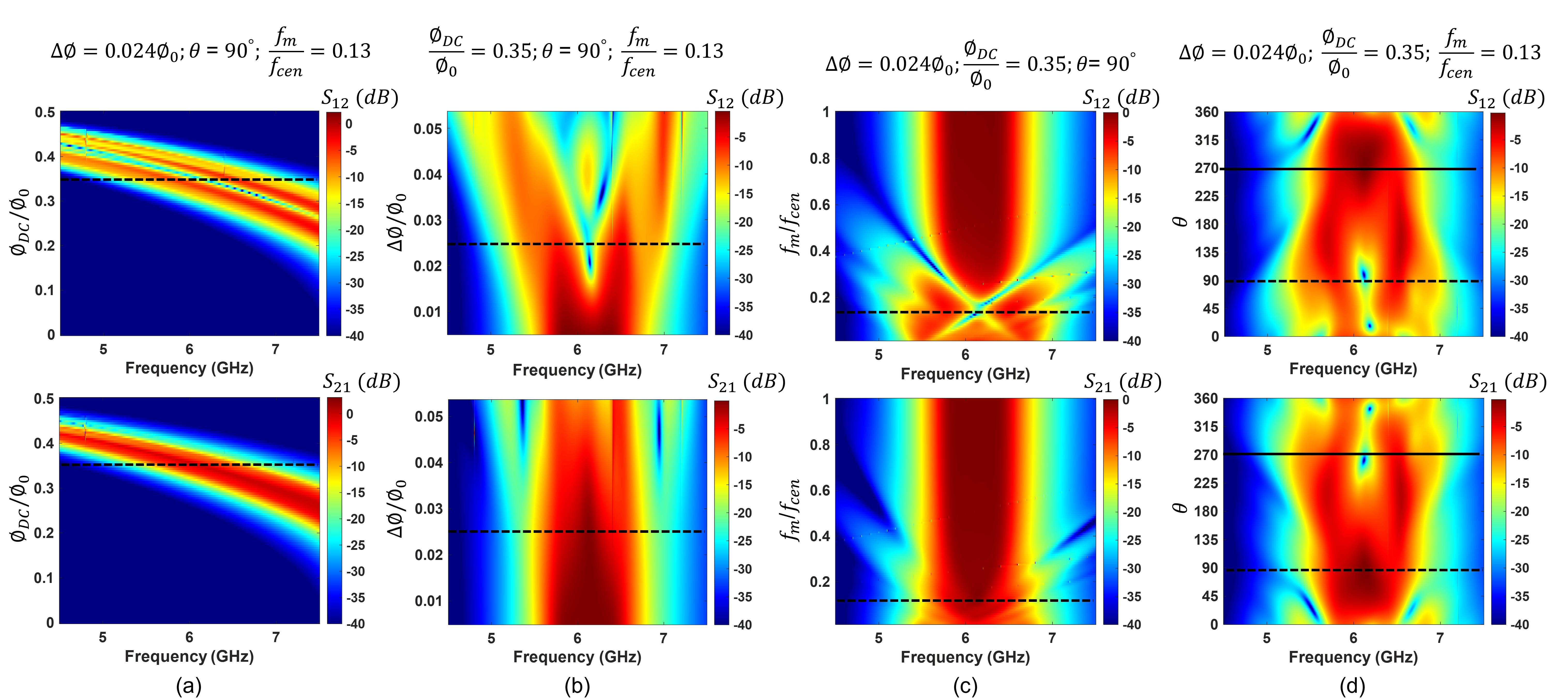}
\caption{\label{fig:BPF_heatmap} Parametric studies of the time-modulated isolator with three coupled resonators. Modulated $S_{12(0,0)}$ and $S_{21(0,0)}$ with varying (a) static flux bias $\Phi_{DC}$, (b) modulation amplitude $\Delta\phi$, (c) modulation frequency $f_m$, and (d) phase staggering $\theta$. Here $S_{21(0,0)} 
  (S_{12(0,0)})$ represents the transmission from port 1 (port 2) to port 2 (port 1) for the same input and output frequencies. The dashed lines represent the optimal biasing for transmission from port 1 to port 2 and isolation from port 2 to port 1. The solid line in panel (d) represents the optimal phase biasing for transmission in the reverse direction.}
\end{figure*}

\subsection{Isolator Using Time-Modulated Series-Coupled Resonators}
Traditional bandpass filters (such as Butterworth, Chebyshev, etc.) are typically implemented by coupling multiple LC resonators using admittance inverters. The number of resonators, their coupling, and loaded quality factor are chosen to achieve the desired filter characteristics, such as bandwidth, out-of-band rejection, and in-band ripple \cite{Pozar_textbook}. Following a similar architecture, we couple multiple unit cells through J-inverters as shown in Fig.~\ref{fig:bpf}(a). Similar to a conventional LC filter, the static flux biased SQUIDs, the parallel capacitors, and the admittance of the J-inverters can be chosen to realize a specific bandpass filter response. When the SQUIDs are biased with a DC Flux, the lack of frequency translation results in a reciprocal transmission through the filter. 

However, modulating the SQUIDs within each resonator with sinusoidal flux pumps that have staggered phase shifts would result in a non-reciprocal transmission response in the circuit. In this scenario, the incident signal is first translated to the intermodulation frequencies ($f_\mathrm{in}\pm kf_\mathrm{m}$) and then reconstructed back to the input frequency. The different phase staggering in the pump signals encountered by the input signal in forward direction leads to constructive addition of the reconstructed IM products in one direction, resulting in low insertion loss. Conversely, in the opposite direction, the reconstructed IM products add up destructively, resulting in high isolation \cite{Gomes_TMTT_2019}. The response of the filter can be analyzed in the same fashion as presented in the section~\ref{sec:amp_NR}. The ABCD matrix of the filter is a cascade of the ABCD of all the admittance inverters and unit cells, that is

\begin{equation}
M_{BPF} = M_{J_1}M_{LC_1}M_{J_2}M_{LC_2} ... M_{J_{n-1}}M_{LC_{n-1}}M_{J_n},
\label{eqn: M_3}
\end{equation}

\noindent where $M_{J_1}$,$M_{J_2}$,$M_{J_{n-1}}$, and $M_{J_n}$ are the spectral-ABCD matrices of the admittance inverters and $M_{LC_1}$,$M_{LC_2}$ and $M_{LC_{n-1}}$ are the spectral-ABCD matrices of the unit cells. Therefore, substituting \eqref{eq:LC_SQUID} and \eqref{eqn: M_J} into \eqref{eqn: M_3} results in the ABCD matrix of the filter. Fig.~\ref{fig:bpf}(b) shows the simulation responses of a second-order bandpass filter optimized for maximum non-reciprocity. However, due to limited degrees-of-freedom and frequency translational paths for converting the intermodulation (IM) products back to the input signal, the second-order BPF exhibits a high insertion loss of 4.64~dB when optimized for an isolation of 20~dB. Increasing the number of resonators within the bandpass filter would result in more degrees of freedom and would enable us to achieve both low insertion loss and high isolation. Therefore, higher-order filters such as third-order and fourth-order filters (shown in Fig.~\ref{fig:bpf}(c) and (d), respectively) can be used to achieve near-zero insertion loss while providing high isolation. Fig.~\ref{fig:bpf}(e) and (f) depict the normalized output spectrum of a third-order filter in the transmission and the isolation directions respectively. Consistent with the scattering parameters, the filter achieves low insertion loss in the forward direction and high isolation in the reverse direction.

In this article, we focused on a detailed analysis of a third-order bandpass filter as it offers low insertion loss, high isolation, and requires less chip area. To maintain consistency with the design implemented in section~\ref{sec:implementation}, we carried out the analytical analysis on a third-order bandpass filter designed to operate at 6~GHz with a bandwidth of 700~MHz. This coupled-resonator isolator consists of four modulation parameters, namely (i) static flux bias $\Phi_{DC}$, (ii) modulation amplitude $\Delta\Phi$, (iii) modulation frequency $f_m$, and (iv) modulation phase staggering $\theta$. Fig.~\ref{fig:BPF_heatmap} depicts the parametric study of the simulated forward and reverse transmission responses across varying modulation parameters, which was performed using the spectral-ABCD matrices. Fig.~\ref{fig:BPF_heatmap}(a) shows the tunability in the center frequency of the isolator across varying static flux bias. Low insertion loss $<$0.5~dB and high isolation $>$20~dB can be achieved across a tuning bandwidth of 1.9~GHz by tuning the static flux bias from 0.3$\Phi_0$ to 0.4$\Phi_0$. Fig.~\ref{fig:BPF_heatmap}(b) depicts the performance across modulation flux amplitude. As expected, when the amplitude decreases to 0, the filter approaches the static design. We found that a flux modulation amplitude of 0.024$\Phi_0$ results in the optimal insertion loss while achieving a high isolation. Fig.~\ref{fig:BPF_heatmap}(c) suggests that when the modulation frequency approaches the bandwidth of the static BPF, it results in the best in isolation. Finally, Fig.~\ref{fig:BPF_heatmap}(d) shows that the filter can achieve non-reciprocity with phase staggering ranging from 40$^{\circ}$ to 120$^{\circ}$. Within this range, higher phase staggering results in higher isolation, however with a reduced transmission bandwidth. We choose to operate with a phase staggering of $90^{\circ}$ as it achieves a low insertion loss of 0.4 dB and a high isolation of $25$ dB. Additionally, note that the transmission and the isolation are off-symmetric with respect to each other when $\theta=\theta+180^{\circ}$. Therefore, the transmission and the isolation directions of the isolator can be easily tuned by changing the phase staggering to $270^{\circ}$. Overall, the coupled-resonator based isolator could results in low insertion loss, high isolation, and tunability in the center frequency and direction of transmission within a compact area on a superconducting chip. A similar architecture achieving the isolator functionality has also been explored in~\cite{beck2022wideband}.

\begin{figure}
\centering
\includegraphics[width=0.6\columnwidth]{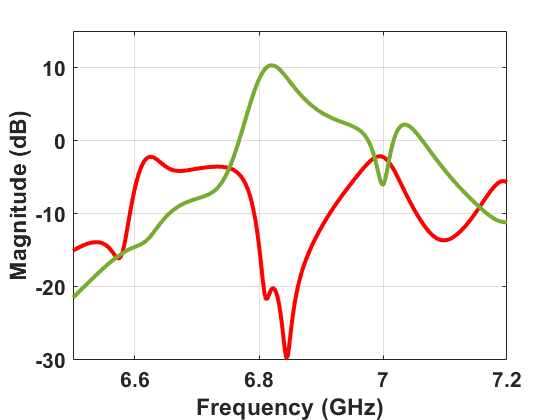}
\caption{\label{fig:Sparam_jpa}Simulated scattering parameters of a series-coupled three resonator network achieving directional amplification by combing the three-wave modulation with the low-frequency phase staggered modulation. }
\end{figure}

\subsection{Uni-Directional Amplification Using Time-Modulated Series-Coupled Resonators}
As discussed earlier, each unit in  the series-coupled isolator is shunted with a SQUID array inductively coupled to a flux line. This flux line can be used to do three-wave mixing which leads to an amplified transmission through the band pass filter. However, this amplification will be bi-directional requiring circulators to protect qubit from the amplified signal. Alternatively, the staggered-phase low-frequency pump modulation can be combined with the three-wave mixing pump to reduce transmission in one direction, hence achieving directional amplification. A simulated result is presented in Fig.~\ref{fig:Sparam_jpa}. We biased the filter at around $6.9$~GHz which corresponds to $0.24 \Phi_{o}$.  A flux tone of $13.8$~GHz has applied to each flux line for three-wave mixing, additionally, a tone of $200$~MHz is applied with phase arrangements such as $0^{o}, 90^{o}$, and $180^{o}$ to each flux line for directionality. The results demonstrate directional amplification, centered around half of the three-wave mixing pump tone, with gains reaching up to 10 dB and isolation up to 20 dB. However, the quantum efficiency of such an amplifier has yet to be experimentally tested. 

\begin{figure*}
\centering
\includegraphics[width=\textwidth]{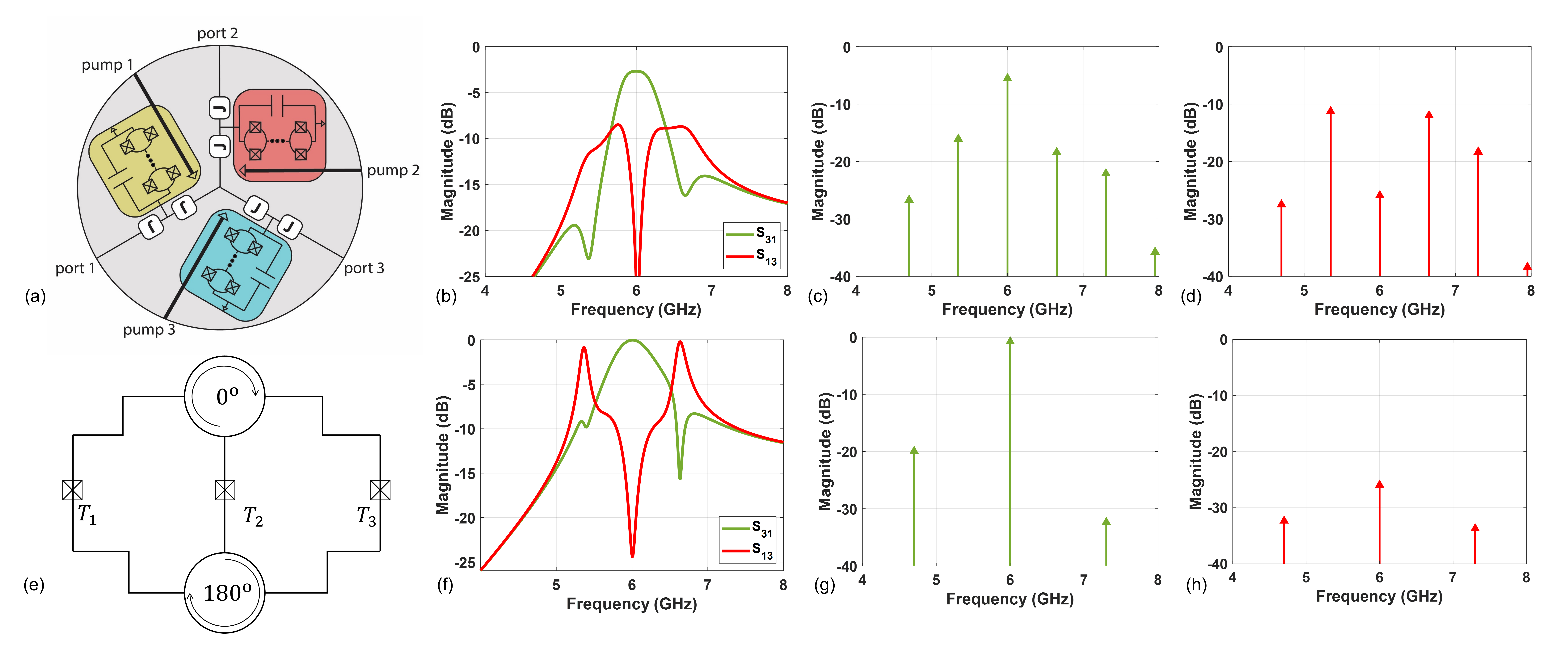}
\caption{\label{fig:Circulator_topology}  Wye-coupled time modulated single-layer circulator (a) topology, (b) scattering parameters, (c) transmission spectrum, and (d) isolation spectrum. Wye-coupled two-layer circulator  (e) topology, (f) scattering parameters, (g) transmission spectrum, and (h) isolation spectrum.}
\end{figure*}

\begin{figure}
\centering
\includegraphics[width=0.6\columnwidth]{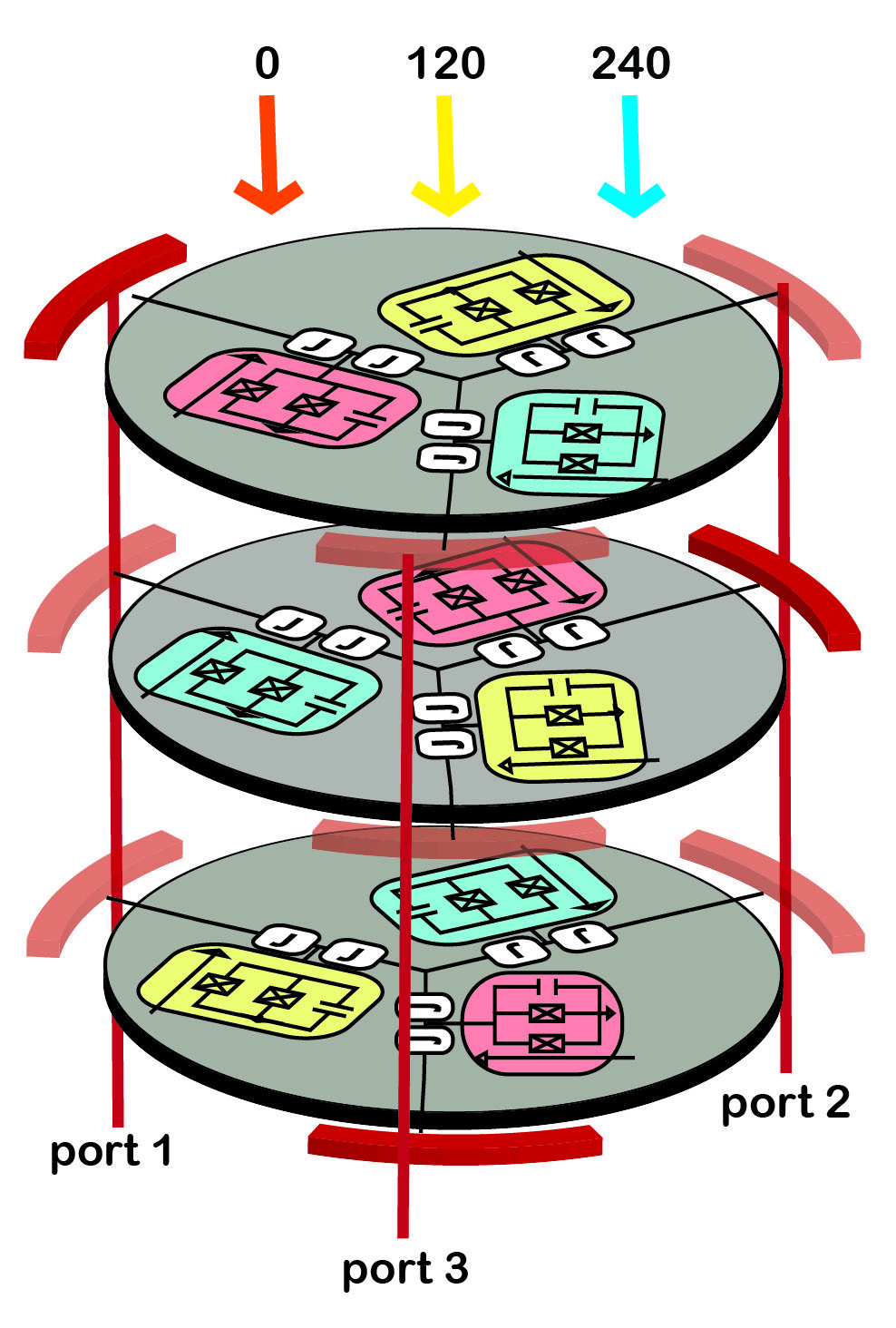}
\caption{\label{fig:3way_circ} Schematic diagram of a 3-port, 3-layer circulator re-using the phase staggered pumps across the layers. }
\end{figure}

\begin{figure*}
\centering
\includegraphics[width=\textwidth]{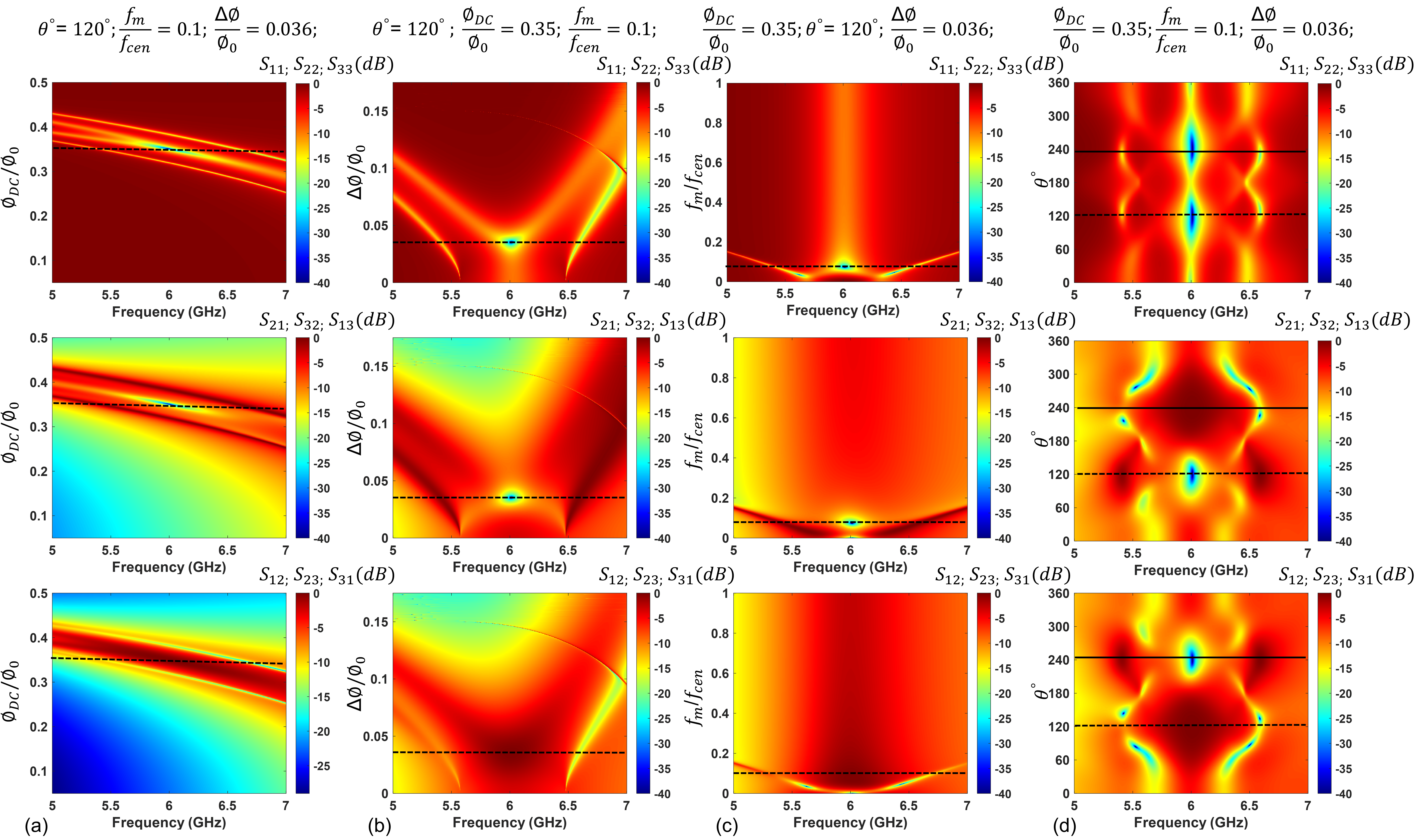}
\caption{\label{fig:Circulator_heatmap}Parametric studies of a three port circulator. Matching, transmission, and isolation across varying (a) static flux bias, (b) modulation flux amplitude, (c) modulation frequency, and (d) phase staggering between the resonators $\theta^{\circ}$. The horizontal dashed lines represent the optimal biasing point for circulation in the clockwise direction.}
\end{figure*}

\subsection{Circulators Using Time-Modulated Wye-Coupled Resonators}
Delta- or wye-topology based circulators can achieve strong non-reciprocity and low insertion losses by coupling time-modulated resonators \cite{Kord_TMTT2018,Alu_ProcIEEE_2020}. In this section, we demonstrate the realization and performance analysis of the circulators with high isolation and low loss using the unit cells presented in section \ref{sec:UnitCell}. As an example, we present the design of a wye-coupled resonator-based circulator. Fig. \ref{fig:Circulator_topology}(a) shows the schematic diagram of a wye-coupled resonator network. With optimal modulation parameters, the circulator can achieve strong non-reciprocity with high isolation ($>$25 dB) and an insertion loss of 2.7 dB, as illustrated in Fig. 
 \ref{fig:Circulator_topology}(b). From the transmission spectrum in the forward direction Fig. \ref{fig:Circulator_topology}(c), it is noticeable that the power is spread out across the IM frequencies, resulting in a relatively high insertion loss.

To eliminate the fundamental harmonic conversion loss from the IM conversion and create a loss-free circulator, another coupled resonator layer can be added in parallel and modulated at a 180$^{\circ}$ phase difference compared to the first layer, as shown in Fig. \ref{fig:Circulator_topology}(e). Fig. \ref{fig:Circulator_topology}(f) depicts the simulated performance of a 2-layer, 3-port circulator achieving near-zero insertion loss with 25 dB isolation. The loss-free transmission and high isolation of the circulator are also apparent in the transmission and isolation spectra depicted in Fig. \ref{fig:Circulator_topology}(g) and (h), respectively. Essentially, the odd IM products ($\omega \pm \omega_m$, $\omega \pm 3\omega_m$) are canceled out due to destructive interference between the 180$^{\circ}$ staggered coupled-resonator layers, leading to low IM products and low insertion loss. Similarly, adding $N$ parallel layers with phase staggering of $\alpha=2\pi/N$ between the layers can cancel out all intermodulation products except those of the form $f_{IM}=(f_{in}\pm k\times Nf_m)$~\cite{Kord_PRA_2019_NWayCirc,Nagulu_JSSC_2021_NWayCirc}. This leads to a spurious-tones-free, loss-free time-modulated circulator, but requires the granularity of several phases within the modulation signals. For a 3-port circulator, it is optimal to design a 3-layer circulator, as the required three modulation phases for the resonators in the first layer can be reconfigured in the second and third layers to achieve the necessary phase staggering without increasing the overhead on the number of pump signals required. A conceptual diagram of a 3-layer, 3-port circulator has been depicted in Fig.~\ref{fig:3way_circ}.

Fig.~\ref{fig:Circulator_heatmap} depict the parametric study of the simulated reflection, transmission, and isolation responses across varying modulation parameters performed using the Floquet scattering matrices~\cite{Cody_TMTT_2020,Tymcheko_TCAS_2021}. 
Fig.~\ref{fig:Circulator_heatmap}(a) shows the tunability in the center frequency of the isolator across varying static flux bias. Low insertion loss $<$0.1 dB and high isolation $>$15~dB can be achieved across a tuning bandwidth of 700 MHz by tuning the static flux bias from 0.339 $\Phi_0$ to 0.368 $\Phi_0$.
Fig.~\ref{fig:Circulator_heatmap}(b) depicts the performance across modulation flux amplitude. We found that a flux modulation amplitude of 0.036$\Phi_0$ results in optimal insertion loss while achieving high isolation.
Fig.~\ref{fig:BPF_heatmap}(c) suggests the optimal modulation frequency is 500 MHz for this design.
Fig.~\ref{fig:Circulator_heatmap}(d) shows the filter can achieve non-reciprocity when phase staggering ranges from 80$^{\circ}$ to 150$^{\circ}$. We choose to operate with a phase staggering of $120^{\circ}$ as it achieves a low insertion loss of 0.02 dB and high isolation of 40 dB. Additionally, note that the transmission and the isolation are off-symmetric with respect to each other when $\theta=\theta+120^{\circ}$. Therefore, the transmission and the isolation directions of the circulator can be easily tuned by changing the phase staggering to $240^{\circ}$.

\begin{figure*}
\centering
\includegraphics[width=\textwidth]{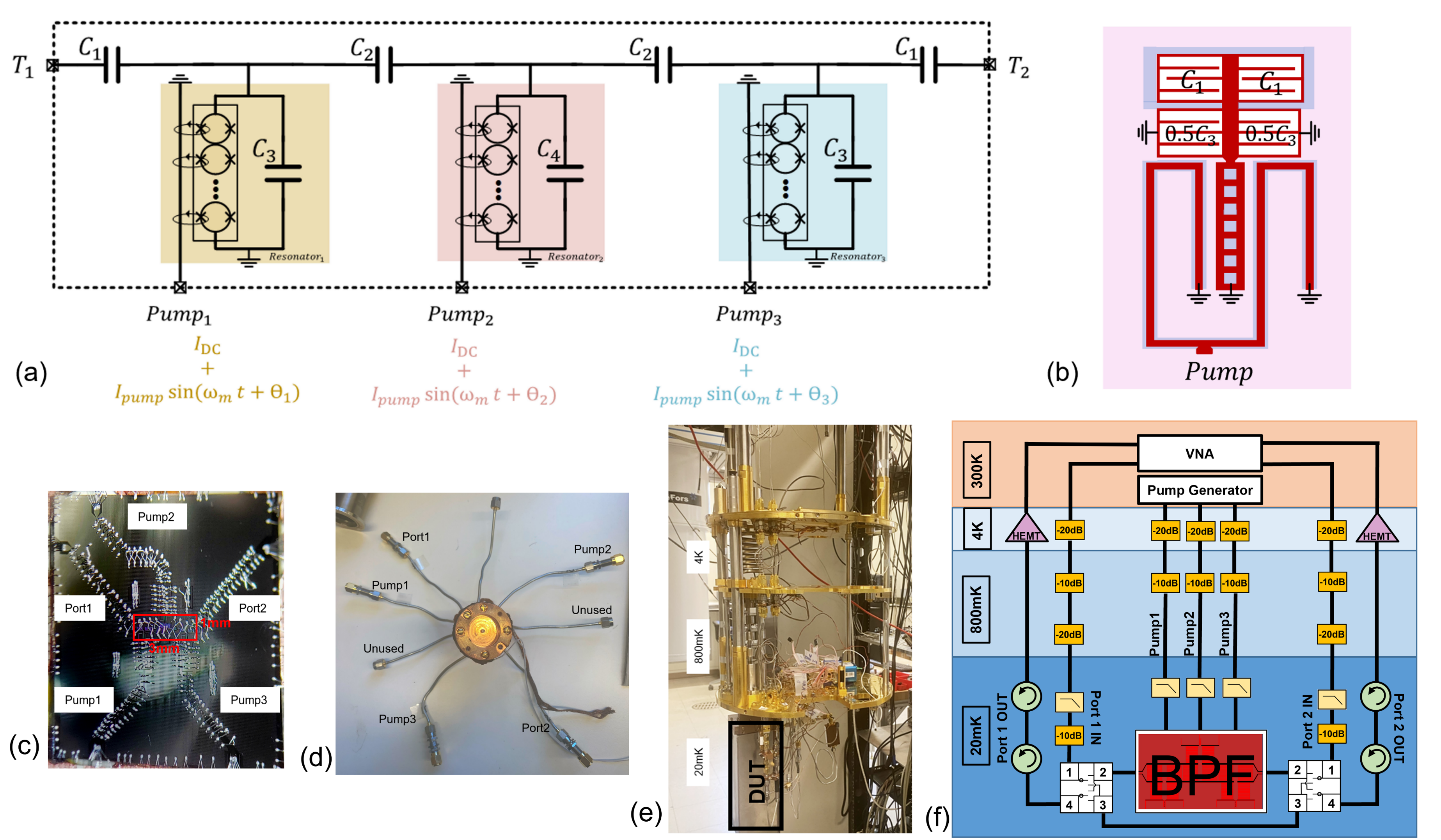}
\caption{\label{fig:bpf_topology} The schematic of a third order bandpass filter implementation using proposed general solution. (b) Layout of the unit cell with capacitive admittance inverters. (c) Copper pacakage to which the chip is wirebonded. (d) Chip microphotograph. (e) Photograph of the dilution refrigerator and (f) measurement block diagram of the experimental setup}
\end{figure*}

\section{Implementation and Measurements of Three Resonator Coupled Isolator}
\label{sec:implementation}

Fig.~\ref{fig:bpf_topology} depicts the schematic of the implemented isolator using three coupled resonators. To achieve a compact chip area, the admittance inverters used for the coupling are realized using $\pi$-capacitive structure. The negative shunt capacitors required within the $\pi$-capacitive are realized by appropriately reducing the shunt capacitor value within the SQUID-resonators. 

\subsection{Design Procedure}
First, a conventional bandpass filter is designed to operate at 6 GHz with a 1 dB bandwidth of 700~MHz and matching $<$-15~dB, and high out-of-band rejection using 10-stacked SQUIDs each with $I_c=4~\mu$A, $\Phi_{DC}=0.35\Phi_o$, and the capacitors  $C_1 = 257$~fF, $C_2 = 113$~fF, $C_3 = 232$~fF and $C_4 = 284$~fF. Then a sinusoidal modulation with progressive phase shifts is imparted to the feed currents to achieve non-reciprocal response, i.e., $\Phi_{feed,i}=\Phi_{DC}+\Delta\Phi\cos(\omega_mt+(i-1)\times\theta)$ where $i$ represents the resonator number from left to right. The filter performance is optimized across the parameter space of $\Delta\Phi, \omega_m, \theta,$ and $\Phi_{DC}$ for low insertion  loss in the forward direction and high isolation in the reverse direction using the parametric study based on spectral-ABCD matrices in Fig.~\ref{fig:BPF_heatmap}. 
For an optimal modulation condition of $\Phi_{DC}=0.35\Phi_0$, $\Delta\Phi$=0.024$\Phi_0, f_m= 700$ MHz, and $\theta=90^{\circ}$, at a center frequency of 6 GHz, the post layout EM simulated filter exhibits an insertion loss of 0.6~dB in the forward direction and isolation $>25$ dB in the reverse direction. Additionally, the center frequency of the non-reciprocal bandpass filter can be tuned by varying the DC flux bias of the SQUID loops from 5.2 GHz to 7.1 GHz while achieving sub-1dB insertion loss and $>$20 dB isolation.

\subsection{Implementation and Fabrication}
The isolator is fabricated using an in-house single-layer aluminum process. The JJs in the filter are realized using the Dolan bridge technique and an electron-beam lithography process on a bi-layer resist stack with an insulator thickness of 2 nm to 5 nm. The required JJ critical current 4~$\mu\mathrm{A}$ is realized using 4~$\mu\mathrm{m} \times$ 0.5~$\mu\mathrm{m}$ junctions with a critical current density of 2~$\mu\mathrm{A}/\mu\mathrm{m}^2$. The SQUID loops in the filter have a conductor width of 4~$\mu\mathrm{m}$ and occupy a total area of $16~\mu\mathrm{m}\times16.5~\mu\mathrm{m}$ with a loop area of $8~\mu\mathrm{m}\times8.5~\mu\mathrm{m}$. The EM simulated geometric loop inductance of the SQUIDs is $12~\mathrm{pH}$, which is approximately  $20\%$ of the SQUID inductance at zero flux bias. The capacitors in the design are created with a conventional fingered-capacitor layout with self-resonance frequency $>$10~GHz. To reduce the capacitive coupling between the feed-line and SQUID loops, a pseudo-differential coupling strategy is employed for the pump line as shown in Fig.~\ref{fig:bpf_topology}(b). The non-reciprocal bandpass filter has a total area of 3 mm$\times$1 mm and the input/output and pump signals are connected to the chip edge using co-planar waveguide launchers.

\begin{figure*}
	\centering
	\includegraphics[width=\textwidth]{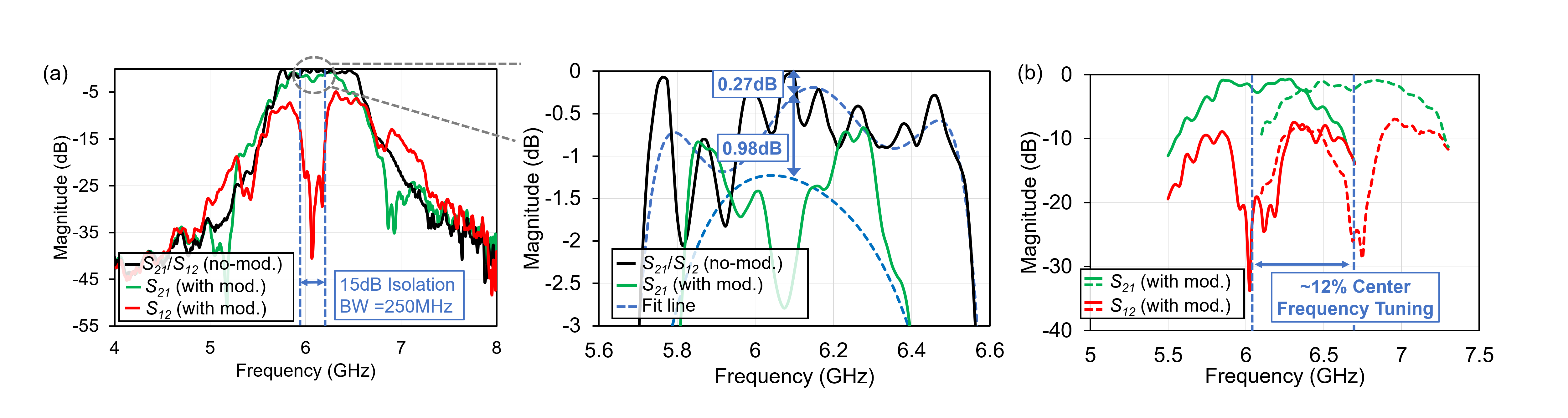}
	\caption{ Measured small signal performance of the coupled-resonator based isolator: (a) transmission when no modulation is applied (black curve), transmission (green curve), and isolation (red curve) when optimal pump-modulation is applied. (b) Measured center frequency tuning by leveraging the DC flux bias of the SQUIDs in the resonators.}
	\label{fig:sps_meas}
\end{figure*} 
\begin{figure*}
	\centering
	\includegraphics[width=\textwidth]{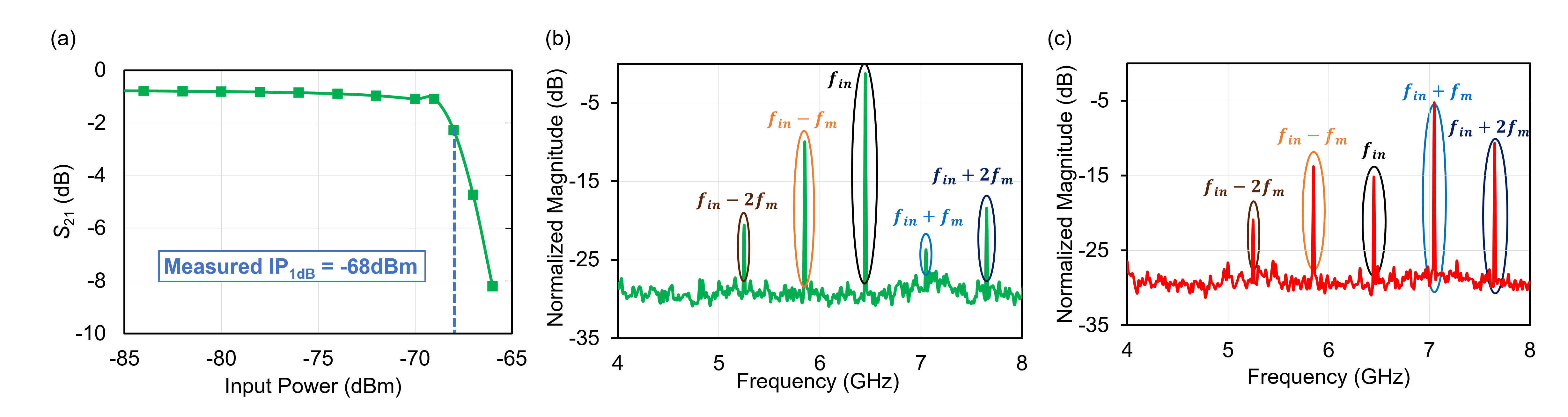}
	\caption{(a) Measured $S_{21}$ at center frequency across varying input power. Measured output spectra normalized to the input power (b) in the forward direction and (c) in the reverse direction.}
	\label{fig:spectrum_meas}
\end{figure*}
\subsection{Measurement Results}
The optical image of the fabricated non-reciprocal bandpass filter is shown in Fig.~\ref{fig:bpf_topology}(c). The device is wire-bonded to a PCB and is mounted in a copper package as shown in Figure~\ref{fig:bpf_topology}(d). The input/output terminals of the bandpass filter and three pump lines are linked to five of the eight RF lines within the copper package. The package is placed inside a dilution refrigerator with a base temperature of 20 mK and surrounded by cryoperm shielding to provide additional protection again external magnetic fields.
Figure~\ref{fig:bpf_topology}(e) and Fig.~\ref{fig:bpf_topology}(f) show a picture and a diagram that outlines the setup used for the measurement. The input/output terminals of the bandpass filter are connected to the probe and readout lines in the dilution refrigerator through two SMA latch transfer switches (model no. Radiall R577432000). A through line is also connected between the remaining ports of the switches in order to measure the transmission loss and/or gain of the probe and readout lines inside the dilution refrigerator.

Figure~\ref{fig:sps_meas}(a) depicts the measured scattering parameters of the non-reciprocal bandpass filter after normalizing with the transmission loss of the through structure. The measured transmission loss of the filter with no modulation is 0.27~dB (black curve) which could be due to the imperfections in the gain-based calibration. When flux-pumping is applied to achieve non-reciprocity, we measured an additional insertion loss of 0.98~dB in the forward direction (green curve). The measured isolation is +25~dB at the center frequency and is $>$15~dB across 250~MHz bandwidth in the reverse direction. Additionally, the DC flux bias can be leveraged to tune the center frequency of the bandpass filter from 6~GHz to 6.75~GHz as shown in Fig.~\ref{fig:sps_meas}(b). A small ripple in the transmission response is noticed in our measurements due to the mismatches within the probe and readout lines. This can be avoided by employing a full 2-port characterization \cite{Leo_SciIns_2013}. Figure~\ref{fig:spectrum_meas}(a) depicts the transmission loss of the filter across varying input powers. We measured the $-1$~dB compression point to be -68~dBm which is several orders of magnitude larger than the typical power used to readout/control the superconducting qubits. Normalized output spectra in the forward and reverse directions are depicted in Fig.~\ref{fig:spectrum_meas}(b) and (c) respectively. As it can be seen, in the transmission direction, most of the input power is concentrated at the input frequency, thus representing low insertion loss. In the reverse direction, however, the input power is translated to intermodulation frequencies resulting in high isolation. Reducing the signal power translation to IM frequencies and consequently reducing the harmonic losses through differential architecture similar to the differential circulators discussed in section-\ref{sec:devices} and similar to~\cite{Alu_IMS_2017} would be an interesting research direction.

\section{Challenges and Future Outlook}
\label{sec:FutureOutlook}

While time-modulated superconducting non-reciprocal components offer several advantages over their ferrite counterparts such as easier system integration, low cost, small size, and easier scalability to large systems, they are not without limitations. One challenge in the current implementations is the generation of multiple modulation signals. Additionally, multiple coaxial cables are interfaced from these devices to 4K/room temperature stages. This issue can be easily circumvented by implementing on-chip phase generation circuitry by leveraging passive superconducting hybrid couplers~\cite{Raafat_IMS_2022} and tunable phase shifters~\cite{Raafat_TMTT_2021}, thus reducing the number of coaxial cables and their associated heat loading. In addition, time-modulated superconducting circulators can also generate spurious emissions at the intermodulation frequencies, which can potentially interfere with  nearby devices. Clever device architectures such as the $N$-layer circulators, coupled with careful design and filtering in the signal path can help mitigate these spurious emissions, but this remains an ongoing challenge in the development of time-modulated superconducting non-reciprocal devices.

As we look to the future, one exciting direction for these components is their extension to superconducting Floquet topological insulator lattices. These lattices can be a new class of quantum devices that can support topologically protected edge states with unique non-reciprocal properties. By exploiting the non-reciprocal behavior of these edge states, superconducting Floquet TI lattices could enable arbitrary signal routing with high fidelity, leading to the development of new, advanced quantum devices. Additionally, by integrating superconducting qubits into the lattices, these devices could also facilitate long-range qubit coupling with reduced decoherence. These devices could also result in dissipation-free non-reciprocal coupling between quantum devices \cite{wang2019non}, enabling scaling to quantum many-body systems where the study of topological edge-states and invariants~\cite{yao18,kuns18} are expected to yield deviations from the paradigmatic bulk-boundary correspondence~\cite{Weidemann2020,Xiao2020}. Recently, significant progress has been made in the development of photonic~\cite{Fan_NatPhoton_2012}, acoustic~\cite{Fleury_TI_2016} and electronic~\cite{Nagulu_NatElectron_2022} Floquet TI lattices, and expanding this knowledge to superconducting lattices could have a transformative impact on the field of quantum computing and related applications. Overall, the extension of superconducting circulators and isolators to Floquet TI lattices represents an exciting direction for future research and development.

\section{Conclusion}
\label{sec:conclusion}

This article presented the concept of time-modulated coupled resonator networks as a means to develop on-chip, magnetless non-reciprocal components for quantum computing systems. We demonstrated that conventional SQUID-based resonators can serve as unit elements and can be combined in various topologies, such as series-coupled resonators, wye-connected and lattice-coupled resonators, to realize a wide range of on-chip non-reciprocal responses, including isolation, circulation, and directional amplification. These coupled-resonator networks provide reconfigurable reciprocal and non-reciprocal responses that are solely dependent on the modulation parameters, such as the static flux bias, modulation amplitude, frequency, and phase staggering. We discussed the design procedure of these non-reciprocal components and evaluated their performance using spectral-ABCD matrices. Finally, we validated our theoretical findings by implementing and measuring an isolator based on three series-coupled resonator networks, thus demonstrating the potential of such networks for future quantum computing systems. By integrating devices in this manner, existing control and readout chains can be simplified, and new control/readout methods can be enabled that are not feasible with commercially available non-reciprocal components.

\section*{Acknowledgements} 

This work is supported by McKelvey Collaboration Initiation Grant (CIG) 2022, and by NSF Grant No. PHY-1752844 (CAREER), and the Air Force Office of Scientific Research (AFOSR) Multidisciplinary University Research Initiative (MURI) Award on Programmable systems with non-Hermitian quantum dynamics (Grant No. FA9550-21-1- 0202) and use of facilities at the Institute of Materials Science and Engineering at Washington University.

\appendix

\section{Spectral-Admittance Matrix of a Time-Varying Inductor}
\label{sec:Methods}
The voltage and current relationship of a time-varying inductor~\cite{Desoer1959,Kurth1977} can be expressed as

\begin{equation}
\label{eq:ind_time_domain}
    i(t) = L(t)^{-1}\int v(t)dt. 
\end{equation}

When the inductance is modulated with a fundamental frequency of $\omega_m$ and input excitation is at $\omega$, the voltage and current would carry the IM products and can be expressed as

\begin{equation}
    i(t) = \sum_{p=-k}^{k}I(\omega+p\omega_m)e^{-j(\omega+p\omega_m) t},
\end{equation}
\begin{equation}
    v(t) = \sum_{p=-k}^{k}V(\omega+p\omega_m)e^{-j(\omega+p\omega_m)t},
\end{equation}

\noindent where $I(\omega+p\omega_m)$ and $V(\omega+p\omega_m)$ are the current and voltage Fourier coefficients of the frequency $(\omega+p\omega_m)$, and $k$ is the farthest IM product that has been calculated. The value of $k$ determines the accuracy of the computation and as $k\rightarrow \infty$ the computation error $\rightarrow 0$. This current and voltage can be expressed in matrix form as
\begin{equation}
    \underline{I}=\begin{bmatrix}
    \vdots \\ I(\omega-2\omega_m) \\ I(\omega-\omega_m) \\ I(\omega)  \\ I(\omega+\omega_m) \\ I(\omega+2\omega_m) \\ \vdots \end{bmatrix},
\end{equation}

\begin{equation}
    \underline{V}=\begin{bmatrix}
    \vdots \\ V(\omega-2\omega_m) \\ V(\omega-\omega_m) \\ V(\omega)  \\ V(\omega+\omega_m) \\ V(\omega+2\omega_m) \\ \vdots \end{bmatrix}.
\end{equation}

Therefore, the integral of voltage can be calculated as
 \begin{equation} 
\int \underline{V(t)}dt = \begin{bmatrix}
 \frac{V(\omega-k\omega_m)}{j(\omega-k\omega_m)} \\ \vdots \\ \frac{V(\omega)}{j\omega}  \\ \vdots \\ \frac{V(\omega+k\omega_m)}{j(\omega+k\omega_m)} \\  \end{bmatrix} =\underline{\Omega}\times\underline{V}, \\
 \end{equation}
where 
 \begin{equation} 
\underline{\Omega} = \begin{bmatrix}
\frac{1}{j(\omega-k\omega_m)} & 0 & 0 &  0 & 0  \\
0 & \ddots&\vdots & \reflectbox{$\ddots$}&0 \\
0 & \cdots  & \frac{1}{j(\omega)}  & \cdots&0 \\
0 & \reflectbox{$\ddots$}&\vdots & \ddots&0 \\
0 & 0 &  0 & 0 & \frac{1}{j(\omega+k\omega_m)}
 \end{bmatrix}.
 \end{equation}
Therefore, \eqref{eq:ind_time_domain} can be expressed in spectral matrix form as
\begin{equation}
\label{eq:admitance}
    \underline{I} = \mathcal{F}(L(t)^{-1})\ast (\underline{\Omega}\times \underline{V}) = \underline{Y_L}\times \underline{V}.
\end{equation}



\begin{thebibliography}{10}
\expandafter\ifx\csname url\endcsname\relax
  \def\url#1{\texttt{#1}}\fi
\expandafter\ifx\csname urlprefix\endcsname\relax\def\urlprefix{URL }\fi
\providecommand{\bibinfo}[2]{#2}
\providecommand{\eprint}[2][]{\url{#2}}

\bibitem{Arute_2019}
\bibinfo{author}{Arute, F.} \emph{et~al.}
\newblock Quantum supremacy using a programmable superconducting processor.
\newblock \emph{\bibinfo{journal}{Nature}} \textbf{\bibinfo{volume}{574}},
  \bibinfo{pages}{505–510} (\bibinfo{year}{2019}).

\bibitem{Bardin_2020}
\bibinfo{author}{Bardin, J.~C.}, \bibinfo{author}{Sank, D.},
  \bibinfo{author}{Naaman, O.} \& \bibinfo{author}{Jeffrey, E.}
\newblock Quantum Computing: An Introduction for Microwave Engineers.
\newblock \emph{\bibinfo{journal}{IEEE Microwave Magazine}}
  \textbf{\bibinfo{volume}{21}}, \bibinfo{pages}{24--44}
  (\bibinfo{year}{2020}).

\bibitem{Bardin_2021}
\bibinfo{author}{Bardin, J.~C.}, \bibinfo{author}{Slichter, D.~H.} \&
  \bibinfo{author}{Reilly, D.~J.}
\newblock Microwaves in Quantum Computing.
\newblock \emph{\bibinfo{journal}{IEEE Journal of Microwaves}}
  \textbf{\bibinfo{volume}{1}}, \bibinfo{pages}{403--427}
  (\bibinfo{year}{2021}).

\bibitem{Josephson_1962}
\bibinfo{author}{{ Josephson}, B.~D.}
\newblock {Possible new effects in superconductive tunneling}.
\newblock \emph{\bibinfo{journal}{Phys. Lett.}} \textbf{\bibinfo{volume}{1}},
  \bibinfo{pages}{251--253} (\bibinfo{year}{1962}).

\bibitem{Rowell_1963}
\bibinfo{author}{{Anderson}, P.~W.} \& \bibinfo{author}{M.{ Rowel}, J.}
\newblock {Probable observation of the Josephson superconducting tunneling
  effect}.
\newblock \emph{\bibinfo{journal}{Phys. Rev. Lett.}}
  \textbf{\bibinfo{volume}{10}}, \bibinfo{pages}{230} (\bibinfo{year}{1963}).

\bibitem{Leo_MicrowMag2019}
\bibinfo{author}{{Ranzani}, L.} \& \bibinfo{author}{{Aumentado}, J.}
\newblock {Circulators at the Quantum Limit: Recent Realizations of
  Quantum-Limited Superconducting Circulators and Related Approaches}.
\newblock \emph{\bibinfo{journal}{IEEE Microwave Magazine}}
  \textbf{\bibinfo{volume}{20}}, \bibinfo{pages}{112--122}
  (\bibinfo{year}{2019}).

\bibitem{Nagulu_NatElectron_2020}
\bibinfo{author}{Nagulu, A.}, \bibinfo{author}{Reiskarimian, N.} \&
  \bibinfo{author}{Krishnaswamy, H.}
\newblock {Non-reciprocal electronics based on temporal modulation}.
\newblock \emph{\bibinfo{journal}{Nature Electronics}}
  \textbf{\bibinfo{volume}{3}}, \bibinfo{pages}{241--250}
  (\bibinfo{year}{2020}).

\bibitem{Alu_ProcIEEE_2020}
\bibinfo{author}{Kord, A.}, \bibinfo{author}{Sounas, D.~L.} \&
  \bibinfo{author}{Alù, A.}
\newblock Microwave Nonreciprocity.
\newblock \emph{\bibinfo{journal}{Proceedings of the IEEE}}
  \textbf{\bibinfo{volume}{108}}, \bibinfo{pages}{1728--1758}
  (\bibinfo{year}{2020}).

\bibitem{Alu_Science2014}
\bibinfo{author}{Fleury, R.} \emph{et~al.}
\newblock {Sound isolation and giant linear nonreciprocity in a compact
  acoustic circulator}.
\newblock \emph{\bibinfo{journal}{Science}} \textbf{\bibinfo{volume}{343}},
  \bibinfo{pages}{516--519} (\bibinfo{year}{2014}).

\bibitem{Fbar_circ_Bhave}
\bibinfo{author}{Torunbalci, M.~M.}, \bibinfo{author}{Odelberg, T.~J.},
  \bibinfo{author}{Sridaran, S.}, \bibinfo{author}{Ruby, R.~C.} \&
  \bibinfo{author}{Bhave, S.~A.}
\newblock {An {FBAR} Circulator}.
\newblock \emph{\bibinfo{journal}{IEEE Microwave and Wireless Components
  Letters}} \textbf{\bibinfo{volume}{28}}, \bibinfo{pages}{395--397}
  (\bibinfo{year}{2018}).

\bibitem{Matteo_MEMS2018}
\bibinfo{author}{Yu, Y.} \emph{et~al.}
\newblock Magnetic-free radio frequency circulator based on spatiotemporal
  commutation of {MEMS} resonators.
\newblock In \emph{\bibinfo{booktitle}{2018 IEEE Micro Electro Mechanical
  Systems (MEMS)}}, \bibinfo{pages}{154--157} (\bibinfo{year}{2018}).

\bibitem{kamal1960}
\bibinfo{author}{Kamal, A.}
\newblock A parametric device as a nonreciprocal element.
\newblock \emph{\bibinfo{journal}{Proceedings of the IRE}}
  \textbf{\bibinfo{volume}{48}}, \bibinfo{pages}{1424--1430}
  (\bibinfo{year}{1960}).

\bibitem{Alu_NatPhys_2014}
\bibinfo{author}{Estep, N.~A.}, \bibinfo{author}{Sounas, D.~L.},
  \bibinfo{author}{Soric, J.} \& \bibinfo{author}{Al{\`u}, A.}
\newblock {Magnetic-free non-reciprocity and isolation based on parametrically
  modulated coupled-resonator loops}.
\newblock \emph{\bibinfo{journal}{Nature Physics}}
  \textbf{\bibinfo{volume}{10}}, \bibinfo{pages}{923--927}
  (\bibinfo{year}{2014}).

\bibitem{NRK_NatComm16}
\bibinfo{author}{Reiskarimian, N.} \& \bibinfo{author}{Krishnaswamy, H.}
\newblock {Magnetic-free non-reciprocity based on staggered commutation}.
\newblock \emph{\bibinfo{journal}{Nat. Commun.}} \textbf{\bibinfo{volume}{7}}
  (\bibinfo{year}{2016}).

\bibitem{TD_NatComm17}
\bibinfo{author}{Dinc, T.} \emph{et~al.}
\newblock Synchronized conductivity modulation to realize broadband lossless
  magnetic-free non-reciprocity.
\newblock In \emph{\bibinfo{booktitle}{Nature Commun.}},
  vol.~\bibinfo{volume}{8} (\bibinfo{year}{2017}).

\bibitem{biedka2017ultra}
\bibinfo{author}{Biedka, M.~M.}, \bibinfo{author}{Zhu, R.},
  \bibinfo{author}{Xu, Q.~M.} \& \bibinfo{author}{Wang, Y.~E.}
\newblock Ultra-wide band non-reciprocity through sequentially-switched delay
  lines.
\newblock \emph{\bibinfo{journal}{Scientific reports}}
  \textbf{\bibinfo{volume}{7}}, \bibinfo{pages}{40014} (\bibinfo{year}{2017}).

\bibitem{Haldane_TI_2008}
\bibinfo{author}{Haldane, F. D.~M.} \& \bibinfo{author}{Raghu, S.}
\newblock Possible Realization of Directional Optical Waveguides in Photonic
  Crystals with Broken Time-Reversal Symmetry.
\newblock \emph{\bibinfo{journal}{Phys. Rev. Lett.}}
  \textbf{\bibinfo{volume}{100}}, \bibinfo{pages}{013904}
  (\bibinfo{year}{2008}).

\bibitem{tzuang2014non}
\bibinfo{author}{Tzuang, L.~D.}, \bibinfo{author}{Fang, K.},
  \bibinfo{author}{Nussenzveig, P.}, \bibinfo{author}{Fan, S.} \&
  \bibinfo{author}{Lipson, M.}
\newblock Non-reciprocal phase shift induced by an effective magnetic flux for
  light.
\newblock \emph{\bibinfo{journal}{Nature photonics}}
  \textbf{\bibinfo{volume}{8}}, \bibinfo{pages}{701--705}
  (\bibinfo{year}{2014}).

\bibitem{chamanara2017optical}
\bibinfo{author}{Chamanara, N.}, \bibinfo{author}{Taravati, S.},
  \bibinfo{author}{Deck-L{\'e}ger, Z.-L.} \& \bibinfo{author}{Caloz, C.}
\newblock {Optical isolation based on space-time engineered asymmetric photonic
  band gaps}.
\newblock \emph{\bibinfo{journal}{Phys. Rev. B}} \textbf{\bibinfo{volume}{96}},
  \bibinfo{pages}{155409} (\bibinfo{year}{2017}).

\bibitem{Kerckhoff2015}
\bibinfo{author}{Kerckhoff, J.}, \bibinfo{author}{Lalumi\`ere, K.},
  \bibinfo{author}{Chapman, B.~J.}, \bibinfo{author}{Blais, A.} \&
  \bibinfo{author}{Lehnert, K.~W.}
\newblock {On-Chip Superconducting Microwave Circulator from Synthetic
  Rotation}.
\newblock \emph{\bibinfo{journal}{Phys. Rev. Applied}}
  \textbf{\bibinfo{volume}{4}}, \bibinfo{pages}{034002} (\bibinfo{year}{2015}).

\bibitem{Lehnert_PRX_2017}
\bibinfo{author}{Chapman, B.~J.} \emph{et~al.}
\newblock {Widely Tunable On-Chip Microwave Circulator for Superconducting
  Quantum Circuits}.
\newblock \emph{\bibinfo{journal}{Phys. Rev. X}} \textbf{\bibinfo{volume}{7}},
  \bibinfo{pages}{041043} (\bibinfo{year}{2017}).

\bibitem{Aumentado_PRA_2017}
\bibinfo{author}{Lecocq, F.} \emph{et~al.}
\newblock {Nonreciprocal Microwave Signal Processing with a Field-Programmable
  Josephson Amplifier}.
\newblock \emph{\bibinfo{journal}{Phys. Rev. Applied}}
  \textbf{\bibinfo{volume}{7}}, \bibinfo{pages}{024028} (\bibinfo{year}{2017}).

\bibitem{Chow_2017}
\bibinfo{author}{{Abdo}, B.}, \bibinfo{author}{{Brink}, M.},  \&
  \bibinfo{author}{{Chow}, J.~M.}
\newblock {Gyrator operation using Josephson mixers}.
\newblock \emph{\bibinfo{journal}{Phys. Rev. Appl.}}
  \textbf{\bibinfo{volume}{8}} (\bibinfo{year}{2017}).

\bibitem{Ranzani_PRA_2017}
\bibinfo{author}{Ranzani, L.} \emph{et~al.}
\newblock Wideband Isolation by Frequency Conversion in a Josephson-Junction
  Transmission Line.
\newblock \emph{\bibinfo{journal}{Phys. Rev. Appl.}}
  \textbf{\bibinfo{volume}{8}}, \bibinfo{pages}{054035} (\bibinfo{year}{2017}).

\bibitem{Stace_PRL_2018}
\bibinfo{author}{Müller, C.}, \bibinfo{author}{Guan, S.},
  \bibinfo{author}{Vogt, N.}, \bibinfo{author}{Cole, J.~H.} \&
  \bibinfo{author}{Stace, T.~M.}
\newblock Passive On-Chip Superconducting Circulator Using a Ring of Tunnel
  Junctions.
\newblock \emph{\bibinfo{journal}{Phys. Rev. Lett.}}
  \textbf{\bibinfo{volume}{120}}, \bibinfo{pages}{213602}
  (\bibinfo{year}{2018}).

\bibitem{Chapman_PRA_2019}
\bibinfo{author}{Chapman, B.~J.}, \bibinfo{author}{Rosenthal, E.~I.} \&
  \bibinfo{author}{Lehnert, K.~W.}
\newblock Design of an On-Chip Superconducting Microwave Circulator with Octave
  Bandwidth.
\newblock \emph{\bibinfo{journal}{Phys. Rev. Applied}}
  \textbf{\bibinfo{volume}{11}}, \bibinfo{pages}{044048}
  (\bibinfo{year}{2019}).

\bibitem{Bretheau_PRR_2021}
\bibinfo{author}{Fatemi, V.}, \bibinfo{author}{Akhmerov, A.~R.} \&
  \bibinfo{author}{Bretheau, L.}
\newblock Weyl Josephson circuits.
\newblock \emph{\bibinfo{journal}{Phys. Rev. Res.}}
  \textbf{\bibinfo{volume}{3}}, \bibinfo{pages}{013288} (\bibinfo{year}{2021}).

\bibitem{Richman_PRXQuantum2021}
\bibinfo{author}{Richman, B.} \& \bibinfo{author}{Taylor, J.~M.}
\newblock Circulation by Microwave-Induced Vortex Transport for Signal
  Isolation.
\newblock \emph{\bibinfo{journal}{PRX Quantum}} \textbf{\bibinfo{volume}{2}},
  \bibinfo{pages}{030309} (\bibinfo{year}{2021}).

\bibitem{Fan_NatPhoton_2012}
\bibinfo{author}{Fang, K.}, \bibinfo{author}{Yu, Z.} \& \bibinfo{author}{Fan,
  S.}
\newblock Realizing effective magnetic field for photons by controlling the
  phase of dynamic modulation.
\newblock \emph{\bibinfo{journal}{Nature Photon.}}
  \textbf{\bibinfo{volume}{6}}, \bibinfo{pages}{782–787}
  (\bibinfo{year}{2012}).

\bibitem{Fleury_TI_2016}
\bibinfo{author}{Fleury, R.}, \bibinfo{author}{Khanikaev, A.~B.} \&
  \bibinfo{author}{Alu, A.}
\newblock Floquet topological insulators for sound.
\newblock \emph{\bibinfo{journal}{Nat. Commun.}} \textbf{\bibinfo{volume}{7}}
  (\bibinfo{year}{2016}).

\bibitem{Nagulu_NatElectron_2022}
\bibinfo{author}{Nagulu, A.} \emph{et~al.}
\newblock Chip-scale Floquet topological insulators for 5G wireless systems.
\newblock \emph{\bibinfo{journal}{Nature Electron.}}
  \textbf{\bibinfo{volume}{5}}, \bibinfo{pages}{300--309}
  (\bibinfo{year}{2022}).

\bibitem{Girvin_PRA_2010}
\bibinfo{author}{Koch, J.}, \bibinfo{author}{Houck, A.~A.},
  \bibinfo{author}{Hur, K.~L.} \& \bibinfo{author}{Girvin, S.~M.}
\newblock Time-reversal-symmetry breaking in circuit-QED-based photon lattices.
\newblock \emph{\bibinfo{journal}{Phys. Rev. A}} \textbf{\bibinfo{volume}{82}},
  \bibinfo{pages}{043811} (\bibinfo{year}{2010}).

\bibitem{Paetznick_PRXQuantum_2023}
\bibinfo{author}{Paetznick, A.} \emph{et~al.}
\newblock Performance of Planar Floquet Codes with Majorana-Based Qubits.
\newblock \emph{\bibinfo{journal}{PRX Quantum}} \textbf{\bibinfo{volume}{4}},
  \bibinfo{pages}{010310} (\bibinfo{year}{2023}).

\bibitem{Alu_IMS_2017}
\bibinfo{author}{Kord, A.}, \bibinfo{author}{Sounas, D.~L.} \&
  \bibinfo{author}{Al{\`u}, A.}
\newblock {Differential magnetless circulator using modulated bandstop
  filters}.
\newblock In \emph{\bibinfo{booktitle}{2017 IEEE MTT-S International Microwave
  Symposium (IMS)}}, \bibinfo{pages}{384--387} (\bibinfo{year}{2017}).

\bibitem{Alejandro_TMTT_2019}
\bibinfo{author}{Wu, X.} \emph{et~al.}
\newblock {Isolating Bandpass Filters Using Time-Modulated Resonators}.
\newblock \emph{\bibinfo{journal}{IEEE Transactions on Microwave Theory and
  Techniques}} \textbf{\bibinfo{volume}{67}}, \bibinfo{pages}{2331--2345}
  (\bibinfo{year}{2019}).

\bibitem{Gomes_TMTT_2019}
\bibinfo{author}{Alvarez-Melcon, A.}, \bibinfo{author}{Wu, X.},
  \bibinfo{author}{Zang, J.}, \bibinfo{author}{Liu, X.} \&
  \bibinfo{author}{Gomez-Diaz, J.~S.}
\newblock {Coupling Matrix Representation of Nonreciprocal Filters Based on
  Time-Modulated Resonators}.
\newblock \emph{\bibinfo{journal}{IEEE Transactions on Microwave Theory and
  Techniques}} \textbf{\bibinfo{volume}{67}}, \bibinfo{pages}{4751--4763}
  (\bibinfo{year}{2019}).

\bibitem{Desoer1959}
\bibinfo{author}{Desoer, C.}
\newblock {Steady-State Transmission through a Network Containing a Single
  Time-Varying Element}.
\newblock \emph{\bibinfo{journal}{IRE Transactions on Circuit Theory}}
  \textbf{\bibinfo{volume}{6}}, \bibinfo{pages}{244--252}
  (\bibinfo{year}{1959}).

\bibitem{Kurth1977}
\bibinfo{author}{Kurth, C.}
\newblock {Steady-state analysis of sinusoidal time-variant networks applied to
  equivalent circuits for transmission networks}.
\newblock \emph{\bibinfo{journal}{IEEE Transactions on Circuits and Systems}}
  \textbf{\bibinfo{volume}{24}}, \bibinfo{pages}{610--624}
  (\bibinfo{year}{1977}).

\bibitem{Pozar_textbook}
\bibinfo{author}{Pozar, D.~M.}
\newblock \emph{\bibinfo{title}{Microwave engineering}}
  (\bibinfo{publisher}{Wiley}, \bibinfo{year}{1998}).

\bibitem{Kord_TMTT2018}
\bibinfo{author}{Kord, A.}, \bibinfo{author}{Sounas, D.~L.} \&
  \bibinfo{author}{Alù, A.}
\newblock Magnet-Less Circulators Based on Spatiotemporal Modulation of
  Bandstop Filters in a Delta Topology.
\newblock \emph{\bibinfo{journal}{IEEE Transactions on Microwave Theory and
  Techniques}} \textbf{\bibinfo{volume}{66}}, \bibinfo{pages}{911--926}
  (\bibinfo{year}{2018}).

\bibitem{Naaman_PRXQuantum2022}
\bibinfo{author}{Naaman, O.} \& \bibinfo{author}{Aumentado, J.}
\newblock Synthesis of Parametrically Coupled Networks.
\newblock \emph{\bibinfo{journal}{PRX Quantum}} \textbf{\bibinfo{volume}{3}},
  \bibinfo{pages}{020201} (\bibinfo{year}{2022}).

\bibitem{beck2022wideband}
\bibinfo{author}{Beck, M.~A.}, \bibinfo{author}{Selvanayagam, M.},
  \bibinfo{author}{Carniol, A.}, \bibinfo{author}{Cairns, S.} \&
  \bibinfo{author}{Mancini, C.~P.}
\newblock Wideband Josephson Parametric Isolator (\bibinfo{year}{2022}).
\newblock \eprint{2212.08563}.

\bibitem{Kord_PRA_2019_NWayCirc}
\bibinfo{author}{Kord, A.}, \bibinfo{author}{Krishnaswamy, H.} \&
  \bibinfo{author}{Al\`u, A.}
\newblock Magnetless Circulators with Harmonic Rejection Based on N-Way
  Cyclic-Symmetric Time-Varying Networks.
\newblock \emph{\bibinfo{journal}{Phys. Rev. Appl.}}
  \textbf{\bibinfo{volume}{12}}, \bibinfo{pages}{024046}
  (\bibinfo{year}{2019}).

\bibitem{Nagulu_JSSC_2021_NWayCirc}
\bibinfo{author}{Nagulu, A.} \emph{et~al.}
\newblock Ultra-Wideband Switched-Capacitor Delays and Circulators—Theory and
  Implementation.
\newblock \emph{\bibinfo{journal}{IEEE Journal of Solid-State Circuits}}
  \textbf{\bibinfo{volume}{56}}, \bibinfo{pages}{1412--1424}
  (\bibinfo{year}{2021}).

\bibitem{Cody_TMTT_2020}
\bibinfo{author}{Scarborough, C.} \& \bibinfo{author}{Grbic, A.}
\newblock Accelerated N-Path Network Analysis Using the Floquet Scattering
  Matrix Method.
\newblock \emph{\bibinfo{journal}{IEEE Transactions on Microwave Theory and
  Techniques}} \textbf{\bibinfo{volume}{68}}, \bibinfo{pages}{1248--1259}
  (\bibinfo{year}{2020}).

\bibitem{Tymcheko_TCAS_2021}
\bibinfo{author}{Tymchenko, M.}, \bibinfo{author}{Nagulu, A.},
  \bibinfo{author}{Krishnaswamy, H.} \& \bibinfo{author}{Alù, A.}
\newblock Universal Frequency-Domain Analysis of N-Path Networks.
\newblock \emph{\bibinfo{journal}{IEEE Transactions on Circuits and Systems I:
  Regular Papers}} \textbf{\bibinfo{volume}{68}}, \bibinfo{pages}{569--580}
  (\bibinfo{year}{2021}).

\bibitem{Leo_SciIns_2013}
\bibinfo{author}{Ranzani, L.}, \bibinfo{author}{Spietz, L.},
  \bibinfo{author}{Popovic, Z.} \& \bibinfo{author}{Aumentado, J.}
\newblock {Two-port microwave calibration at millikelvin temperatures}.
\newblock \emph{\bibinfo{journal}{Review of Scientific Instruments}}
  \textbf{\bibinfo{volume}{84}}, \bibinfo{pages}{034704}
  (\bibinfo{year}{2013}).

\bibitem{Raafat_IMS_2022}
\bibinfo{author}{Khaira, N.~K.}, \bibinfo{author}{Singh, T.} \&
  \bibinfo{author}{Mansour, R.~R.}
\newblock {Cryogenic Wideband Quadrature Hybrid Couplers Implemented in a Low
  Temperature Superconductor Multilayer Process}.
\newblock In \emph{\bibinfo{booktitle}{2022 IEEE/MTT-S International Microwave
  Symposium - IMS 2022}}, \bibinfo{pages}{160--163} (\bibinfo{year}{2022}).

\bibitem{Raafat_TMTT_2021}
\bibinfo{author}{Singh, T.}, \bibinfo{author}{Khaira, N.~K.} \&
  \bibinfo{author}{Mansour, R.~R.}
\newblock {Thermally Actuated SOI RF MEMS-Based Fully Integrated Passive
  Reflective-Type Analog Phase Shifter for mmWave Applications}.
\newblock \emph{\bibinfo{journal}{IEEE Transactions on Microwave Theory and
  Techniques}} \textbf{\bibinfo{volume}{69}}, \bibinfo{pages}{119--131}
  (\bibinfo{year}{2021}).

\bibitem{wang2019non}
\bibinfo{author}{Wang, Y.-X.} \& \bibinfo{author}{Clerk, A.}
\newblock Non-Hermitian dynamics without dissipation in quantum systems.
\newblock \emph{\bibinfo{journal}{Physical Review A}}
  \textbf{\bibinfo{volume}{99}}, \bibinfo{pages}{063834}
  (\bibinfo{year}{2019}).

\bibitem{yao18}
\bibinfo{author}{Yao, S.} \& \bibinfo{author}{Wang, Z.}
\newblock Edge States and Topological Invariants of Non-Hermitian Systems.
\newblock \emph{\bibinfo{journal}{Phys. Rev. Lett.}}
  \textbf{\bibinfo{volume}{121}}, \bibinfo{pages}{086803}
  (\bibinfo{year}{2018}).

\bibitem{kuns18}
\bibinfo{author}{Kunst, F.~K.}, \bibinfo{author}{Edvardsson, E.},
  \bibinfo{author}{Budich, J.~C.} \& \bibinfo{author}{Bergholtz, E.~J.}
\newblock Biorthogonal Bulk-Boundary Correspondence in Non-Hermitian Systems.
\newblock \emph{\bibinfo{journal}{Phys. Rev. Lett.}}
  \textbf{\bibinfo{volume}{121}}, \bibinfo{pages}{026808}
  (\bibinfo{year}{2018}).

\bibitem{Weidemann2020}
\bibinfo{author}{Weidemann, S.} \emph{et~al.}
\newblock Topological funneling of light.
\newblock \emph{\bibinfo{journal}{Science}} \textbf{\bibinfo{volume}{368}},
  \bibinfo{pages}{311--314} (\bibinfo{year}{2020}).

\bibitem{Xiao2020}
\bibinfo{author}{Xiao, L.} \emph{et~al.}
\newblock Non-Hermitian bulk{\textendash}boundary correspondence in quantum
  dynamics.
\newblock \emph{\bibinfo{journal}{Nature Physics}}
  \textbf{\bibinfo{volume}{16}}, \bibinfo{pages}{761--766}
  (\bibinfo{year}{2020}).

\end{thebibliography}

\end{document}